\documentclass[11pt,leqno]{article}

\usepackage{multirow}
\usepackage{eepic,epic}
\usepackage{epsfig}
\usepackage{graphicx}
\usepackage{color}
\usepackage{mathptmx}
\usepackage{amssymb}
\usepackage{amsmath}
\usepackage{pifont}
\usepackage{nicefrac}
\input epsf %
\usepackage{algorithmic}

\newcommand{\OMIT}[1]{} %
\newtheorem{theorem}{Theorem}[section]

\newtheorem{corollary}[theorem]{Corollary}
\newtheorem{definition}[theorem]{Definition}
\newtheorem{lemma}[theorem]{Lemma}

\newtheorem{example}[theorem]{Example}

\newenvironment{proofs}{\noindent{\bf Proof.}\hspace*{1em}}{\literalqed\bigskip}
\def\literalqed{{\ \nolinebreak\hfill\mbox{\qedblob\quad}}}

\newcommand\qedblob{\mbox{\ding{113}}}

\newcommand{\rationals}{\ensuremath{{ \mathbb{Q} }}}

\newcommand{\rationalszeroone}{\ensuremath{{ \mathbb{Q}_{[0,1]} }}}
\newcommand{\matrixrationalszeroone}[2]{\ensuremath{{ \mathbb{Q}_{[0,1]}^{#1{\times }#2} }}}

\newcommand{\wtwo}{{\ensuremath{{\rm{W}}[2]}}}
\newcommand{\wone}{{\ensuremath{{\rm{W}}[1]}}}
\newcommand{\wt}[1]{{\ensuremath{{\rm{W}}[#1]}}}

\newcommand{\p}{\ensuremath{\rm P}}

\newcommand{\np}{{\ensuremath{\rm NP}}}

\newcommand{\dtime}{{\ensuremath{\rm DTIME}}}

\newcommand{\naturalnumber}{\ensuremath{{  \mathbb{N} }}}
\def\nats{\naturalnumber}

\newcommand{\condition}{\,\mid \:}

\newcommand{\MAX}{\textsc{Max}}
\newcommand{\MIN}{\textsc{Min}}
\newcommand{\ol}{\textsc{OL}}
\newcommand{\owl}{\textsc{OWL}}
\newcommand{\plp}[2]{\textsc{{#1}-{#2}-PLP}}
\newcommand{\plpiw}[2]{\textsc{{#1}-{#2}-PLP-WIW}}

\newcommand{\YES}{\textup{\textsf{YES}}}
\newcommand{\NO}{\textup{\textsf{NO}}}
 \newcommand{\Oh}{{\mathcal{O}}}
\newcommand{\FPT}{\ensuremath{\rm FPT}}

\newcommand{\problemsize}[1]{|#1|}

\begin{document}

\title{The Complexity of Probabilistic Lobbying\thanks{A preliminary
version of this paper appears in the proceedings of the \emph{1st
International Conference on Algorithmic Decision Theory}, October
2009 \cite{erd-fer-gol-mat-rai-rot:c:probabilistic-lobbying}.
This work was supported in part by DFG grants
RO~\mbox{1202/11-1} and RO~\mbox{1202/12-1}, the European Science
Foundation's EUROCORES program LogICCC, the Alexander von Humboldt
Foundation's TransCoop program,
and by NSF grants EAGER-CCF-1049360 and ITR-0325063.  This work
was done in part while the second author was affiliated to 
Heinrich-Heine-Universit\"{a}t D\"{u}sseldorf and visiting Universit\"at
Trier, while the first and the fourth author were visiting
Heinrich-Heine-Universit\"at D\"usseldorf, and while the sixth author
was visiting the University of Rochester.}}

\author{
Daniel 
Binkele-Raible,\thanks{URL: www.informatik.uni-trier.de/\symbol{126}raible.
Universit\"{a}t Trier, FB 4---Abteilung Informatik, %
54286 Trier,
Germany.
}
\and
G\'{a}bor Erd\'{e}lyi,\thanks{URL: 
ccc.cs.uni-duesseldorf.de/\symbol{126}erdelyi.
Nanyang Technological University,
Division of Mathematical Science,
Singapore 639798.
}
\and
Henning Fernau,\thanks{URL: www.informatik.uni-trier.de/\symbol{126}fernau.
Universit\"{a}t Trier, FB 4---Abteilung Informatik, %
54286 Trier,
Germany.
}
\and
Judy Goldsmith,\thanks{URL: www.cs.uky.edu/\symbol{126}goldsmit.
University of Kentucky,
Dept.\ of Computer Science,
Lexington, KY 40506, 
USA.
}
\and
Nicholas Mattei,\thanks{URL: www.cs.uky.edu/\symbol{126}nsmatt2.
University of Kentucky,
Dept.\ of Computer Science,
Lexington, KY 40506, 
USA.
}
\and
J\"{o}rg Rothe\thanks{URL: ccc.cs.uni-duesseldorf.de/\symbol{126}rothe.
Heinrich-Heine-Universit\"{a}t D\"{u}sseldorf,
Institut f\"{u}r Informatik, 
40225 D\"{u}sseldorf,
Germany.
} 
}

\date{February 21, 2011}

\maketitle

\begin{abstract}
We propose  models for lobbying in a probabilistic environment,
in which an actor (called ``The Lobby'') seeks to influence 
voters' preferences of voting for or against multiple issues when the
voters' preferences are represented in terms of probabilities.  In
particular, we provide two evaluation criteria and two bribery
methods to formally describe these models, and we consider the
resulting forms of lobbying with and without issue weighting.  We
provide a formal analysis for these problems of lobbying in a
stochastic environment, and determine their classical and
parameterized complexity depending on the given bribery/evaluation
criteria and on various natural parameterizations.
Specifically, we show that some of these problems can be
solved in polynomial time, some are $\np$-complete but fixed-parameter
tractable, and some are $\wtwo$-complete.  Finally, we provide
approximability and inapproximability results for these problems and
 several variants.
\\[2mm]
\end{abstract}

\section{Introduction}

\subsection{Motivation and Informal Description of Probabilistic Lobbying Models}
 \label{sec:introduction-motivation}

In most democratic political systems, laws are passed by elected officials
who are supposed to represent their constituency.  Individual entities
such as citizens or corporations are not supposed
to have undue influence in the wording or passage of a law.  However,
they are allowed to make contributions to representatives, and it is
common to include an indication that the contribution carries an
expectation
that
the representative will vote a certain way on a particular issue.

Many factors can affect a representative's vote on a particular issue.
There are the representative's personal beliefs about the issue, which
presumably were part of the reason that the constituency elected them.
There are also the campaign contributions, communications from
constituents, communications from potential donors, and the
representative's own expectations of further contributions and
political support.

It is a complicated process to reason about.  Earlier work considered
the problem of meting out contributions to representatives in order to
pass a set of laws or influence a set of votes.  However, the earlier
computational complexity work on this problem made the assumption that
a politician who accepts a contribution will in fact---if the
contribution meets a given threshold---vote according to the wishes of
the donor.

It is said that ``An honest politician is one who stays bought,'' but
that does not take into account the ongoing pressures from personal
convictions and opposing lobbyists and donors.  We consider the
problem of influencing a set of votes under the assumption that we
can influence only the \emph{probability} that the politician votes
as we desire.

There are several axes along which we complicate the picture.  The first is
the notion of sufficiency:  What does it mean to say we have donated
enough to influence the vote?  Does it mean that the probability that
a single vote will go our way is greater than some threshold?  That the
probability that all the votes go our way is greater than that
threshold?  We formally define and 
discuss these and other criteria in the section
on evaluation criteria (Section~\ref{sec:eval}).  In particular,
we consider two methods for evaluating the outcome of a vote:
\begin{enumerate}
\item \emph{strict majority}, where a vote on an issue is won by a
  strict majority of voters having a probability of accepting this
  issue that exceeds a given threshold, and

\item \emph{average majority}, where a vote on an issue is won exactly
  when the voters' average probability of accepting this issue exceeds
  a given threshold.

\end{enumerate}

How does one donate money to a campaign?  In the United States there
are several laws that influence how, when, and how much a particular
person or organization can donate to a particular candidate.  We
examine ways in which money can be channeled into the political
process in the section on bribery methods
(Section~\ref{sec:briberymethods}).  In particular,
we consider two methods that an actor (called ``The Lobby'') can use
to influence the voters' preferences of voting for or against multiple 
issues:
\begin{enumerate}
\item \emph{microbribery}, where The Lobby may choose which voter to
  bribe on which issue in order to influence the outcome of the vote
  according to the evaluation criterion used and

\item \emph{voter bribery}, where The Lobby may choose which voters to
  bribe and for each voter bribed the funds are equally distributed
  over all the issues, again aiming at changing the outcome of the
  vote according to the evaluation criterion used.
\end{enumerate}

The voter bribery method is due to
Christian et al.~\cite{chr-fel-ros-sli:j:lobbying},
who were the first to study lobbying in the context of direct
democracy where voters vote on multiple referenda.  Their ``Optimal
Lobbying'' problem (denoted $\ol$) is a deterministic and unweighted
variant of the lobbying problems that we present in this paper.
We state this problem in the standard format for parameterized complexity:

\begin{description}
\item[Name:] \textsc{Optimal Lobbying}.
\item[Given:] An $m {\times} n$ $0/1$ matrix~$E$ and a $0/1$ vector
$\vec{Z}$ of length~$n$.  Each row of $E$ represents a voter and each
column represents an issue.  An entry of $E$ is $1$ if this voter
votes ``yes'' for this issue, and is $0$ otherwise.
$\vec{Z}$ represents The Lobby's target outcome.
\item[Parameter:] A positive integer $b$
(representing the number of voters to be influenced).
\item[Question:] Is there a choice of $b$ rows of the matrix (i.e., of
$b$ voters) that can be changed such that in each column of the
resulting matrix (i.e., for each issue) a strict majority vote yields the
outcome targeted by The Lobby?
\end{description}

Christian et al.~\cite{chr-fel-ros-sli:j:lobbying}
proved that $\ol$ is $\wtwo$-complete.
Sandholm noted that the ``Optimal Weighted Lobbying'' ($\owl$)
problem, which allows different voters to have different prices and so
generalizes $\ol$, can be expressed as and solved via the ``binary
multi-unit combinatorial reverse auction winner-determination
problem''
(see~\cite{san-sur-gil-lev:c:combinatorial-auction-generalizations}).

The microbribery method in the context of lobbying---though inspired
by the different notion of microbribery that
Faliszewski et al.~\cite{fal-hem-hem-rot:c:llull,fal-hem-hem-rot:c:copeland-fully-resists-constructive-control,fal-hem-hem-rot:j:llull-copeland-full-techreport}
introduced in the context of bribery in voting---is new to this paper.

\subsection{Organization of This Paper and a Brief Overview of Results}

Christian et al.~\cite{chr-fel-ros-sli:j:lobbying} show that $\ol$
is complete for the (parameterized) complexity
class $\wtwo$.
We extend their model of lobbying as mentioned above (see
Section~\ref{sec:models-for-probabilistic-lobbying} for a formal
description), and provide algorithms and analysis for these extended
models in terms of classical and parameterized complexity.
All complexity-theoretic notions needed will be presented in
Section~\ref{sec:background-parameterized-complexity}.

Our classical complexity results (presented in
Section~\ref{sec:classical-complexity}) are shown via either
polynomial-time algorithms for or reductions showing
$\np$-completeness of the problems studied.  For the parameterized
complexity results (presented in
Section~\ref{sec:parameterized-complexity}), we choose natural
parameters such as The Lobby's budget, the budget per referendum, and
the ``discretization level'' used in formalizing our probabilistic
lobbying problems (see Section~\ref{sec:initialmodel}).  We also
consider the concept of issue weighting, modeling that certain issues
will be of more importance to The Lobby than others.  Our classical
and parameterized complexity results are summarized in
Table~\ref{tab:plp} (see page~\pageref{tab:plp}) for problems without
and in Table~\ref{tab:plpiw} (see page~\pageref{tab:plpiw}) for
problems with issue weighting.

In Section~\ref{sec:approximability}, we provide approximability and
inapproximability results for probabilistic lobbying problems.  In
this way we add breadth and depth to not only the models but also the
understanding of lobbying behavior.  We conclude by
summarizing our main results and stating some open problems in
Section~\ref {sec:conclusions}.

\subsection{Related Work}

Lobbying has been studied formally by economists, computer scientists,
and special interest groups since at least 1983
\cite{rei:j:formal-lobby} and as an extension to formal game theory
since 1944 \cite{neu-mor:b:games}.  The different disciplines have considered
mostly disjoint aspects of the process while seeking to accomplish
distinct goals with their respective formal models.  Economists study lobbying
 as ``economic games,'' as
defined by
von Neumann and Morgenstern~\cite{neu-mor:b:games}.  This
analysis is focused on learning how these complex systems work and
deducing optimal strategies for winning the competitions
\cite{rei:j:formal-lobby,bay-kov-dev:j:rig,bay-kov-dev:j:completeinfo}.
This work has also focused on how to ``rig'' a vote and how to optimally
dispense the funds among the various individuals~\cite{bay-kov-dev:j:rig}.
Economists are interested in finding
effective and efficient bribery schemes \cite{bay-kov-dev:j:rig} as
well as determining strategies for instances of two or more players
\cite{bay-kov-dev:j:rig,rei:j:formal-lobby,bay-kov-dev:j:completeinfo}.
Generally, they reduce the problem of finding an effective lobbying
strategy to one of finding a winning strategy for the specific type of
game.
Economists have also formalized this problem for bribery systems in
both the United States \cite{rei:j:formal-lobby} and the European
Union~\cite{cro:j:eulobby}.

The study of lobbying from a computational perspective that was
initiated by
Christian et al.~\cite{chr-fel-ros-sli:j:lobbying} falls into the
emerging field of computational social choice, which stimulates 
a bidirectional transfer between social choice theory (in particular,
voting and preference aggregation) and computer science.  For example,
voting systems have been applied in various areas of artificial
intelligence, most notably in the design of multiagent systems (see,
e.g., \cite{eph-ros:j:multiagent-planning}), for developing
recommender systems \cite{gho-mun-her-sen:c:voting-for-movies}, for
designing a meta-search engine that aggregates the website rankings
generated by several search engines
\cite{dwo-kum-nao-siv:c:rank-aggregation}, etc. Applications of voting
systems in such automated settings (not restricted only to political
elections in human societies) requires a better understanding of the
computational properties of the problems related to voting.  In
particular, many papers have focused on the complexity of
\begin{itemize}
\item \emph{winner determination} (see, e.g.,
  \cite{bar-tov-tri:j:who-won,hem-hem-rot:j:dodgson,rot-spa-vog:j:young,hem-spa-vog:j:kemeny}),

\item \emph{manipulation}
  (see, e.g., \cite{bar-tov-tri:j:manipulating,bar-orl:j:polsci:strategic-voting,con-san:c:voting-tweaks,elk-lip:c:polsci:universal-tweaks-coalitions,con-san-lan:j:when-hard-to-manipulate,hem-hem:j:dichotomy-scoring,pro-ros:j:juntas,bre-fal-hem-sch-sch:c:approximabillity-of-manipulating-elections,mei-pro-ros-zoh:j:multiwinner,zuc-pro-ros:c-With-Ptr-to-toappear-journal-version:coalitional-manipulation,fal-hem-hem:j:bribery,fal-hem-hem-rot:c:single-peaked-preferences}),

\item \emph{procedural control} (see, e.g.,
  \cite{bar-tov-tri:j:control,hem-hem-rot:j:destructive-control,fal-hem-hem-rot:j:llull-copeland-full-techreport,hem-hem-rot:j:hybrid,fal-hem-hem-rot:c:single-peaked-preferences,erd-now-rot:j:sp-av,erd-rot:c:fallback-voting,erd-pir-rot:c-toappear:voter-partition-in-bucklin-and-fallback-voting}),
  and

\item \emph{bribery in elections} (see, e.g.,
  \cite{fal-hem-hem:j:bribery,fal-hem-hem-rot:j:llull-copeland-full-techreport}).
\end{itemize}
For more details, the reader is referred to the surveys
by Faliszewski et al.~\cite{fal-hem-hem-rot:b:richer} and 
Baumeister et 
al.~\cite{bau-erd-hem-hem-rot:b:computational-apects-of-approval-voting}
and the references cited therein.  In comparison, much less work has
been done on \emph{lobbying in voting on multiple referenda} that we
are concerned with here (\cite{chr-fel-ros-sli:j:lobbying}, see also
\cite{san-sur-gil-lev:c:combinatorial-auction-generalizations}).

\section{Models for Probabilistic Lobbying}
\label{sec:models-for-probabilistic-lobbying}

\subsection{Initial Model}
\label{sec:initialmodel}

We begin with a simplistic version of the \textsc{Probabilistic
  Lobbying Problem} (\textsc{PLP}, for short), in which voters start
with initial probabilities of voting for an issue and are assigned
known costs for increasing their probabilities of voting according to
``The Lobby's'' agenda by each of a finite set of increments.
The question, for this class of problems, is: Given the above 
information, along with an agenda and a 
fixed budget $B$, can The Lobby target its bribes in order 
to achieve its agenda?

The complexity of 
the problem seems to hinge on the evaluation criterion for 
what it means to ``win a vote'' or 
``achieve an agenda.''  We discuss the possible 
interpretations of evaluation and bribery later
in this section.\footnote{We
stress that when we use the term ``bribery'' in this paper, it is
meant in the sense of lobbying~\cite{chr-fel-ros-sli:j:lobbying},
not in the sense
Faliszewski et al.~\cite{fal-hem-hem:j:bribery} have in mind when
defining
bribery in elections (see also, e.g.,
\cite{fal-hem-hem-rot:c:llull,fal-hem-hem-rot:c:copeland-fully-resists-constructive-control,fal-hem-hem-rot:j:llull-copeland-full-techreport}).}
First, however, we will formalize the problem
by defining data objects needed to represent the problem
instances.  (A similar model was first 
discussed by Reinganum~\cite{rei:j:formal-lobby} in the continuous 
case and we translate it here to the discrete case.  
This will allow us to present algorithms for, and a complexity 
analysis of, the problem.)

Let $\matrixrationalszeroone{m}{n}$ denote the set of
$m {\times} n$ matrices over $\rationals_{[0,1]}$ (the
rational numbers in the interval $[0,1]$).
We say $P \in \matrixrationalszeroone{m}{n}$ is a \emph{probability 
matrix (of size $m {\times} n$)},
where each entry $p_{i,j}$ of $P$ gives the probability that voter
$v_{i}$ will vote ``yes'' for referendum (synonymously, for issue)
$r_{j}$.  The result of a vote is either a ``yes'' (represented by
$1$) or a ``no'' (represented by $0$).  Thus, we represent the
result of any vote on all issues as a $0/1$ vector $\vec{X} = (x_1,
x_2, \dots, x_n)$, which is sometimes also denoted as a string in
$\{0,1\}^n$.

We now associate with each voter/issue pair $(v_i,r_j)$
a discrete price function $c_{i,j}$ for changing $v_i$'s
probability of voting ``yes'' for issue~$r_j$.
Intuitively, $c_{i,j}$
gives the cost for The Lobby of raising or
lowering (in discrete steps) the $i$th voter's probability of voting
``yes'' on the $j$th issue.  A formal description is as follows.

Given the entries $p_{i,j} = \nicefrac{a_{i,j}}{b_{i,j}}$ of a
probability matrix $P \in \matrixrationalszeroone{m}{n}$, where
$a_{i,j} \in \nats = \{0, 1, \ldots\}$ and
$b_{i,j} \in \nats_{>0} = \{1, 2, \ldots\}$, choose some
$k \in \nats$ such that $k+1$ is a common multiple of all $b_{i,j}$,
where $1 \leq i \leq m$ and $1 \leq j \leq n$, and partition the
probability interval $[0,1]$ into $k+1$ steps of size
$\nicefrac{1}{(k+1)}$ each.\footnote{There is some arbitrariness in
  this choice of~$k$.  One might think of more flexible ways of
  partitioning $[0,1]$.  We have chosen this way for the sake of
  simplifying the representation, but we mention that all that matters
  is that for each $i$ and~$j$,
  the discrete price function $c_{i,j}$
  is defined on the value $p_{i,j}$, and is set to zero for this value.}
The integer $k$ will be called the \emph{discretization level} of the
problem instance,
 and each integer $\kappa$, $0\leq \kappa\leq k+1$ might be called a \emph{(confidence) step}.
For each $i \in \{1, 2, \ldots , m\}$ and $j \in \{1, 2, \ldots ,
n\}$, $c_{i,j} : \{0,\nicefrac{1}{(k+1)},\nicefrac{2}{(k+1)}, \ldots
,\nicefrac{k}{(k+1)}, 1\} \rightarrow \nats$ is the
\emph{(discrete) price function for $p_{i,j}$}, i.e.,
$c_{i,j}(\nicefrac{\ell}{(k+1)})$ is the price for changing the
probability of the $i$th voter voting ``yes'' on the $j$th issue from
$p_{i,j}$ to $\nicefrac{\ell}{(k+1)}$, where $0 \leq \ell \leq k+1$.
Note that the domain of $c_{i,j}$ consists of $k+2$ elements of
$\rationalszeroone$ including~$0$, $p_{i,j}$, and~$1$.  In particular,
we require $c_{i,j}(p_{i,j})=0$, i.e., a cost of zero is associated
with leaving the initial probability of voter $v_i$ voting on issue
$r_j$ unchanged.  Note that $k=0$ means $p_{i,j} \in \{0,1\}$, i.e.,
in this case each voter either accepts or rejects each issue with
certainty and The Lobby can only flip these results.\footnote{This is
  the special case of
$\ol$~\cite{chr-fel-ros-sli:j:lobbying}.}
The image of $c_{i,j}$ consists of $k+2$ nonnegative integers
including~$0$ (the confidence steps), and we require that,
for any two elements $a,b$ in the domain of $c_{i,j}$, if
$p_{i,j} \leq a \leq b$ or $p_{i,j} \geq a \geq b$, then
$c_{i,j}(a) \leq c_{i,j}(b)$.  This guarantees monotonicity on the prices in both directions.

We represent the list of
price functions associated with
a probability matrix $P$ as a table $C_P$, called \emph{cost matrix}
in the following,  whose $m \cdot n$ rows give the
price functions $c_{i,j}$ and whose $k+2$
columns give the costs $c_{i,j}(\nicefrac{\ell}{(k+1)})$, 
where $0 \leq \ell \leq k+1$.
Occasionally, we use $c_{i,j}[\ell]$ to denote
$c_{i,j}(\nicefrac{\ell}{(k+1)})$, for $\ell \in \{0, 1, \ldots, k+1\}$.
Note that we choose the same $k$ for each $c_{i,j}$, so we have
the same number of columns in each row of~$C_P$.
The entries of $C_P$ can be
thought of as ``price tags'' indicating what The Lobby must 
pay in order to change the probabilities of voting.

The Lobby also has an integer-valued budget $B$ and an ``agenda,''
which we will denote as a vector $\vec{Z} \in \{0,1\}^{n}$ for $n$
issues, containing the outcomes The Lobby would like
to see on these issues.  For The Lobby, the prices for a bribery that moves
the outcomes of a referendum into the wrong direction do not matter.
Hence, if $\vec{Z}$ is zero at position $j$, then we set
$c_{i,j}(a)=--$ (indicating an unimportant entry)
for $a>p_{i,j}$, and  if $\vec{Z}$ is one at position $j$, then we set
$c_{i,j}(a)=--$ (indicating an unimportant entry)
for $a<p_{i,j}$. Without loss of generality,
we may also assume that $c_{i,j}(a)=0$ if and only if
$a=p_{i,j}$.

For simplicity, we may
assume
that The Lobby's agenda is all ``yes'' votes, so the target vector is
$\vec{Z} = 1^{n}$.  This assumption can be made without loss of
generality, since if there is a zero in $\vec{Z}$ at position~$j$, we
can flip this zero to one and also change the corresponding
probabilities $p_{1,j}, p_{2,j}, \dots , p_{m,j}$ in the $j$th column
of $P$ to $1-p_{1,j}, 1-p_{2,j}, \dots , 1-p_{m,j}$. (See the
evaluation criteria in Section~\ref{sec:eval} for how to determine the
result of voting on a referendum.)
Moreover, the rows of the cost matrix $C_P$ that correspond to 
issue $j$ have to be mirrored.

\begin{example}
Consider the following problem instance with $k=9$ (so there are
$k+1 = 10$ steps), $m=2$ voters, and
$n = 3$ issues.  We will use this as a running example for
the rest of this paper.  In addition to the above definitions for~$k$,
$m$, and~$n$, we also give the following probability matrix $P$ and
cost matrix $C_P$ for~$P$.  (Note that
this example is normalized for an agenda of $\vec{Z} = 1^3$, which is
why The Lobby has no incentive for lowering the acceptance
probabilities, so those costs are omitted below.)

Our example consists of
a probability matrix~$P$:
\[
\begin{array}{|c||c|c|c|}
\hline
    & r_1 & r_2 & r_3 \\ \hline\hline
v_1 & 0.8 & 0.3 & 0.5 \\ \hline
v_2 & 0.4 & 0.7 & 0.4 \\ \hline
\end{array}
\]
and the corresponding cost matrix~$C_P$:
{\small
\[
\begin{array}{|@{\hspace*{1mm}}l@{\hspace*{1mm}}||@{}r@{\hspace*{1mm}}|@{}r@{\hspace*{1mm}}|@{}r@{\hspace*{1mm}}|@{}r@{\hspace*{1mm}}|@{}r@{\hspace*{1mm}}|@{}r@{\hspace*{1mm}}|@{}r@{\hspace*{1mm}}|@{}r@{\hspace*{1mm}}|@{}r@{\hspace*{1mm}}|@{}r@{\hspace*{1mm}}|@{}r@{\hspace*{1mm}}|@{}}
\hline
\multicolumn{1}{@{}|@{}c@{\hspace*{1mm}}||}{c_{i,j}} &
\multicolumn{1}{c@{\hspace*{1mm}}|}{0.0} &
\multicolumn{1}{c@{\hspace*{1mm}}|}{0.1} &
\multicolumn{1}{c@{\hspace*{1mm}}|}{0.2} &
\multicolumn{1}{c@{\hspace*{1mm}}|}{0.3} &
\multicolumn{1}{c@{\hspace*{1mm}}|}{0.4} &
\multicolumn{1}{c@{\hspace*{1mm}}|}{0.5} &
\multicolumn{1}{c@{\hspace*{1mm}}|}{0.6} &
\multicolumn{1}{c@{\hspace*{1mm}}|}{0.7} &
\multicolumn{1}{c@{\hspace*{1mm}}|}{0.8} &
\multicolumn{1}{c@{\hspace*{1mm}}|}{0.9} &
\multicolumn{1}{c@{\hspace*{1mm}}|}{1.0} \\
\hline\hline
c_{1,1} & -- & -- &   -- &  -- &  -- & -- &  -- & -- & 0 & 100 & 140
 \\ \hline
c_{1,2} & -- & -- & -- & 0 & 10 & 70 &  100 &   140 &  310 &  520 & 600
 \\ \hline
c_{1,3} & -- & -- & -- &  -- &  -- &  0 &  15 &  25 &  70 &  90 & 150
 \\ \hline
c_{2,1} & -- & -- & -- &  -- &  0 & 30 &  40 &  70 &  120 & 200 & 270
 \\ \hline
c_{2,2} &  -- &  -- &  -- &  -- &  -- & -- & -- & 0 & 10 & 40 & 90
 \\ \hline
c_{2,3} & -- & -- & -- & -- &   0 & 70 &  90 & 100 & 180 & 300 & 450
 \\ \hline
\end{array}
\]
}
\label{ex:runex}
\end{example}

In Section~\ref{sec:briberymethods}, we describe two bribery
methods, i.e., two specific ways 
in which The Lobby can influence the voters.  These
will be referred to as
\emph{microbribery} (\textsc{MB})
and
\emph{voter bribery} (\textsc{VB}).  
In Section~\ref{sec:eval}, we define two ways in which The Lobby can win
a set of votes.
These evaluation criteria will be referred to as
\emph{strict majority} (\textsc{SM}) and
\emph{average majority} (\textsc{AM}). %
The four basic probabilistic lobbying problems we will study (each a
combination of \textsc{MB}/\textsc{VB} bribery under
\textsc{SM}/\textsc{AM} %
evaluation) are defined in
Section~\ref{sec:basic-plp}, and a modification of these basic
problems with additional issue weighting is introduced in
Section~\ref{sec:issueweighting}.

\subsection{Bribery Methods}
\label{sec:briberymethods}
We begin by first formalizing the bribery methods by which 
The Lobby can influence votes on issues.  We will define two
methods for donating this money.  %

\subsubsection{Microbribery (MB)}

The first method at the disposal of The Lobby
is what we will call \emph{microbribery}.\footnote{Although our
  notion was inspired by theirs, we stress that it should not be
  confused with the term ``microbribery'' used by
Faliszewski et al.~\cite{fal-hem-hem-rot:c:llull,fal-hem-hem-rot:c:copeland-fully-resists-constructive-control,fal-hem-hem-rot:j:llull-copeland-full-techreport}
  in the different context of bribing ``irrational'' voters in
  Llull/Copeland elections via flipping single entries in their
  preference tables.}  We define microbribery to be the editing of
individual elements of the $P$ matrix according to the costs in the
$C_P$ matrix.  Thus The Lobby picks not only which voter
to influence but also which issue to influence for that voter.  This
bribery method allows a very flexible version of bribery, and models
private donations made to politicians or voters in support of specific issues.

More formally, if voter $i$ is bribed with $d$ dollars on issue $j$,
then all entries $c_{i,j}[\ell]$, $0\leq \ell\leq k+1$, of $C_P$
are updated as follows:

$$c_{i,j}[\ell]:=\left\{\begin{array}{ll}
 -- & \text{if} (c_{i,j}[\ell]=--) \lor ((c_{i,j}[\ell]-d)\leq 0)\\
c_{i,j}[\ell]-d & \text{if} (c_{i,j}[\ell]-d)> 0.\\
\end{array}
\right.
$$
Moreover, assuming The Lobby's target vector is $1^n$,
let $T=\operatorname{argmax}\{\ell\mid c_{i,j}[\ell]=--\}$.
Replace $c_{i,j}[T]:=0$ and update $p_{i,j}:=\nicefrac{T}{(k+1)}$.

\begin{example}[continuing Example~\ref{ex:runex}]
\label{exa:runex-100dollars}
To make this concrete, reconsidering our Example~\ref{ex:runex}:
Suppose we give \$100 to the second voter and ask her to change
her opinion on the third issue. This would lead to the following 
update of $C_P$:

{\small
\[
\begin{array}{|@{\hspace*{1mm}}l@{\hspace*{1mm}}||@{}r@{\hspace*{1mm}}|@{}r@{\hspace*{1mm}}|@{}r@{\hspace*{1mm}}|@{}r@{\hspace*{1mm}}|@{}r@{\hspace*{1mm}}|@{}r@{\hspace*{1mm}}|@{}r@{\hspace*{1mm}}|@{}r@{\hspace*{1mm}}|@{}r@{\hspace*{1mm}}|@{}r@{\hspace*{1mm}}|@{}r@{\hspace*{1mm}}|@{}}
\hline
\multicolumn{1}{@{}|@{}c@{\hspace*{1mm}}||}{c_{i,j}} &
\multicolumn{1}{c@{\hspace*{1mm}}|}{0.0} &
\multicolumn{1}{c@{\hspace*{1mm}}|}{0.1} &
\multicolumn{1}{c@{\hspace*{1mm}}|}{0.2} &
\multicolumn{1}{c@{\hspace*{1mm}}|}{0.3} &
\multicolumn{1}{c@{\hspace*{1mm}}|}{0.4} &
\multicolumn{1}{c@{\hspace*{1mm}}|}{0.5} &
\multicolumn{1}{c@{\hspace*{1mm}}|}{0.6} &
\multicolumn{1}{c@{\hspace*{1mm}}|}{0.7} &
\multicolumn{1}{c@{\hspace*{1mm}}|}{0.8} &
\multicolumn{1}{c@{\hspace*{1mm}}|}{0.9} &
\multicolumn{1}{c@{\hspace*{1mm}}|}{1.0} \\
\hline\hline
c_{1,1} & -- & -- &   -- &  -- &  -- & -- &  -- & -- & 0 & 100 & 140
 \\ \hline
c_{1,2} & -- & -- & -- & 0 & 10 & 70 &  100 &   140 &  310 &  520 & 600
 \\ \hline
c_{1,3} & -- & -- & -- &  -- &  -- &  0 &  15 &  25 &  70 &  90 & 150
 \\ \hline
c_{2,1} & -- & -- & -- &  -- &  0 & 30 &  40 &  70 &  120 & 200 & 270
 \\ \hline
c_{2,2} &  -- &  -- &  -- &  -- &  -- & -- & -- & 0 & 10 & 40 & 90
 \\ \hline
c_{2,3} & -- & -- & -- & -- &   -- & -- &  -- & 0 & 80 & 200 & 350
 \\ \hline
\end{array}
\]
}

Accordingly, the matrix $P$ is updated as follows:

\[
\begin{array}{|c||c|c|c|}
\hline
    & r_1 & r_2 & r_3 \\ \hline\hline
v_1 & 0.8 & 0.3 & 0.5 \\ \hline
v_2 & 0.4 & 0.7 & 0.7 \\ \hline
\end{array}
\]
\end{example}

\subsubsection{Voter Bribery (VB)}

The
second method at the disposal of The Lobby
is \emph{voter bribery}.  We can see from the $P$
matrix that each row represents what an individual voter thinks about
all the issues on the docket.  In this method of bribery, The Lobby
picks a voter and then pays to edit the entire row at once with the
funds being equally distributed over all the issues.  So, for
$d$ dollars a fraction of $\nicefrac{d}{n}$
is spent on each issue, and the probabilities change
accordingly.  The cost of moving the voter is given by
the $C_P$ matrix as before.  This method of bribery is analogous to
``buying'' or pushing a single politician or voter.  The Lobby seeks
to donate so much money to some individual voters that they have no
choice but to move all of their votes toward The Lobby's agenda.

Let us be more precise.
To avoid problems with fractions of dollars, we will
assume that the bribery money is donated in multiples of $n$, the number of issues.
Hence, whole dollars will be donated per referendum.
So, if we bribe voter $i$ by giving her $x\cdot n$ dollars,
this results in microbribing every issue by giving $x$ dollars
to voter $i$ to raise her confidence on issue $j$; in other words,
all the entries of $c_{i,j}$ (in~$C_P$), $1\leq j\leq n$, will be
edited (and accordingly~$P$).

\begin{example}[continuing Example~\ref{ex:runex}]
\label{exa:runex-300dollars}
  Let us reconsider our running example: What happens if we give
  \$$300$ to the second voter?  The update of $C_P$ would be as
  follows:

{\small
\[
\begin{array}{|@{\hspace*{1mm}}l@{\hspace*{1mm}}||@{}r@{\hspace*{1mm}}|@{}r@{\hspace*{1mm}}|@{}r@{\hspace*{1mm}}|@{}r@{\hspace*{1mm}}|@{}r@{\hspace*{1mm}}|@{}r@{\hspace*{1mm}}|@{}r@{\hspace*{1mm}}|@{}r@{\hspace*{1mm}}|@{}r@{\hspace*{1mm}}|@{}r@{\hspace*{1mm}}|@{}r@{\hspace*{1mm}}|@{}}
\hline
\multicolumn{1}{@{}|@{}c@{\hspace*{1mm}}||}{c_{i,j}} &
\multicolumn{1}{c@{\hspace*{1mm}}|}{0.0} &
\multicolumn{1}{c@{\hspace*{1mm}}|}{0.1} &
\multicolumn{1}{c@{\hspace*{1mm}}|}{0.2} &
\multicolumn{1}{c@{\hspace*{1mm}}|}{0.3} &
\multicolumn{1}{c@{\hspace*{1mm}}|}{0.4} &
\multicolumn{1}{c@{\hspace*{1mm}}|}{0.5} &
\multicolumn{1}{c@{\hspace*{1mm}}|}{0.6} &
\multicolumn{1}{c@{\hspace*{1mm}}|}{0.7} &
\multicolumn{1}{c@{\hspace*{1mm}}|}{0.8} &
\multicolumn{1}{c@{\hspace*{1mm}}|}{0.9} &
\multicolumn{1}{c@{\hspace*{1mm}}|}{1.0} \\
\hline\hline
c_{1,1} & -- & -- &   -- &  -- &  -- & -- &  -- & -- & 0 & 100 & 140
 \\ \hline
c_{1,2} & -- & -- & -- & 0 & 10 & 70 &  100 &   140 &  310 &  520 & 600
 \\ \hline
c_{1,3} & -- & -- & -- &  -- &  -- &  0 &  15 &  25 &  70 &  90 & 150
 \\ \hline
c_{2,1} & -- & -- & -- &  -- &  -- & -- &  -- &  0 &  20 & 100 & 170
 \\ \hline
c_{2,2} &  -- &  -- &  -- &  -- &  -- & -- & -- & -- & -- & -- & 0
 \\ \hline
c_{2,3} & -- & -- & -- & -- &   -- & -- &  -- & 0 & 80 & 200 & 350
 \\ \hline
\end{array}
\]
}

Accordingly, the matrix $P$ is updated as follows:

\[
\begin{array}{|c||c|c|c|}
\hline
    & r_1 & r_2 & r_3 \\ \hline\hline
v_1 & 0.8 & 0.3 & 0.5 \\ \hline
v_2 & 0.7 & 1.0 & 0.7 \\ \hline
\end{array}
\]
\end{example}

\subsubsection*{Simple Observation}

Note that
microbribery is equivalent to voter bribery if
there is only one referendum.

\subsection{Evaluation Criteria}
\label{sec:eval}
Defining criteria for how an issue is won is the next important step in 
formalizing our models.  Here we define two methods
that one could use to evaluate the eventual outcome of a vote.  
Since we are focusing on problems that are probabilistic in nature,
it is important to 
note that no evaluation criterion will guarantee a win.  
The criteria below yield different 
outcomes depending on the model and problem instance.

\subsubsection{Strict Majority (SM)}

For each issue, a strict majority of the individual voters have
probability greater than some threshold, $t$, of voting according to
the agenda.  In our running example (see Example~\ref{ex:runex}), with
$t=50\%$, the result of the votes would be $\vec{X} = (0,0,0)$,
because none of the issues has a strict majority of voters with above
$50\%$ likelihood of voting ``yes.''  The bribery action described in
Example~\ref{exa:runex-100dollars} results in the same vector,
$(0,0,0)$.  However, the bribery action described in
Example~\ref{exa:runex-300dollars} results in the vector $(1,0,0)$.

\subsubsection{Average Majority (AM)}

For each issue $r_{j}$ of a given probability matrix~$P$, we define
the average probability
$\overline{p_{j}} = \nicefrac{\left(\sum^{m}_{i=1} p_{i,j}\right)}{m}$
of voting ``yes'' for~$r_j$.
We now evaluate the vote to say that $r_{j}$ is 
accepted if and only if $\overline{p_{j}} > t$ where $t$ 
is some threshold.  
In our running example with $t = 50\%$, this would
give us a result vector of $\vec{X} = (1,0,0)$.  However, the bribery
action described in Example~\ref{exa:runex-100dollars} results in the
vector $(1,0,1)$, while the bribery action described in
Example~\ref{exa:runex-300dollars} results in the vector $(1,1,1)$.

\subsubsection*{Simple Observation}

Note that the first two criteria coincide if there is only one voter or if the
discretization level equals zero.

\subsection{Basic Probabilistic Lobbying Problems}
\label{sec:basic-plp}

We now introduce the
four basic problems that we will study.
Recalling that, without loss of generality, The Lobby's target vector
may assumed to be all ones, we define the following problem for
$\textsc{X} \in \{\textsc{MB},
\textsc{VB}\}$ and
$\textsc{Y} \in \{\textsc{SM}, \textsc{AM}\}$.

\begin{description}
\item[Name:] \textsc{X-Y Probabilistic Lobbying Problem}.
\item[Given:] A probability matrix 
$P \in \matrixrationalszeroone{m}{n}$ with 
a cost matrix $C_P$ (with integer entries),
a budget~$B$, and some threshold $t\in \mathbb{Q}_{[0,1[}$.
\item[Question:] Is there a way for The Lobby to influence $C_P$
and hence $P$ (using bribery method
\textsc{X} and evaluation criterion
\textsc{Y},
without exceeding budget $B$) such that the result of the votes on all
issues equals~$1^n$?
\end{description}

We abbreviate this problem name as $\plp{X}{Y}$.

\begin{example}[continuing Example~\ref{ex:runex}]
\label{exa:runex-basic-PLP}
Consider voter bribery and the average majority criterion with our
running example and suppose The Lobby has a budget of $\$75$, i.e., our
instance of $\plp{\textsc{VB}}{\textsc{AM}}$ is $(P,C_P,75)$ with $P$
and $C_P$ as given in Example~\ref{ex:runex}.  Giving $\$75$ to the
first voter would suffice to lift all issues above the threshold of
$50\%$ on average according to the wishes of The Lobby.  The updated
cost matrix~$C'_P$ would be: {\small
\[
\begin{array}{|@{\hspace*{1mm}}l@{\hspace*{1mm}}||@{}r@{\hspace*{1mm}}|@{}r@{\hspace*{1mm}}|@{}r@{\hspace*{1mm}}|@{}r@{\hspace*{1mm}}|@{}r@{\hspace*{1mm}}|@{}r@{\hspace*{1mm}}|@{}r@{\hspace*{1mm}}|@{}r@{\hspace*{1mm}}|@{}r@{\hspace*{1mm}}|@{}r@{\hspace*{1mm}}|@{}r@{\hspace*{1mm}}|@{}}
\hline
\multicolumn{1}{@{}|@{}c@{\hspace*{1mm}}||}{c_{i,j}} &
\multicolumn{1}{c@{\hspace*{1mm}}|}{0.0} &
\multicolumn{1}{c@{\hspace*{1mm}}|}{0.1} &
\multicolumn{1}{c@{\hspace*{1mm}}|}{0.2} &
\multicolumn{1}{c@{\hspace*{1mm}}|}{0.3} &
\multicolumn{1}{c@{\hspace*{1mm}}|}{0.4} &
\multicolumn{1}{c@{\hspace*{1mm}}|}{0.5} &
\multicolumn{1}{c@{\hspace*{1mm}}|}{0.6} &
\multicolumn{1}{c@{\hspace*{1mm}}|}{0.7} &
\multicolumn{1}{c@{\hspace*{1mm}}|}{0.8} &
\multicolumn{1}{c@{\hspace*{1mm}}|}{0.9} &
\multicolumn{1}{c@{\hspace*{1mm}}|}{1.0} \\
\hline\hline
c_{1,1} & -- & -- &   -- &  -- &  -- & -- &  -- & -- & 0 & 75 & 115
 \\ \hline
c_{1,2} & -- & -- & -- & -- & 0 & 45 &  75 &   115 &  285 &  495 & 575
 \\ \hline
c_{1,3} & -- & -- & -- &  -- &  -- &  -- &  -- &  0 &  45 &  65 & 125
 \\ \hline
c_{2,1} & -- & -- & -- &  -- &  0 & 30 &  40 &  70 &  120 & 200 & 270
 \\ \hline
c_{2,2} &  -- &  -- &  -- &  -- &  -- & -- & -- & 0 & 10 & 40 & 90
 \\ \hline
c_{2,3} & -- & -- & -- & -- &   0 & 70 &  90 & 100 & 180 & 300 & 450
 \\ \hline
\end{array}
\]
}
This leads to the following updated probability matrix~$P'$, 
enriched with the average probabilities:
\[
\begin{array}{|c||c|c|c|}
\hline
    & r_1 & r_2 & r_3 \\ \hline\hline
v_1 & 0.8 & 0.4 & 0.7 \\ \hline
v_2 & 0.4 & 0.7 & 0.4 \\ \hline\hline
\overline{p_j} & 0.6 & 0.55 & 0.55 \\\hline
\end{array}
\]
Since each referendum passes the evaluation test, as desired by The Lobby, $(P,C_P,75) \in
\plp{\textsc{VB}}{\textsc{AM}}$.
\end{example}

Notice that the discretization level is an implicit (unary) parameter
of the problem that is indirectly specified through the cost matrix~$C_P$.

\subsection{Probabilistic Lobbying with Issue Weighting}
\label{sec:issueweighting}

We now augment the model to include the concept of issue 
weighting.  It is reasonable to surmise that certain issues will
be of more importance to The Lobby than others.  For this 
reason we will allow The Lobby to specify higher weights to the 
issues that they deem more important.  These positive integer weights 
will be defined for each issue.

We will specify these weights as a vector
$\vec{W} \in \nats_{>0}^{n}$
with size $n$ equal to the total number of issues in our problem
instance.  The higher the weight, the more important that particular
issue is to The Lobby.
Along with the weights for each issue we are also given an objective
value $O \in \nats_{>0}$, which is the minimum weight The Lobby
wants to see passed.
Since this is a partial ordering, it is possible
for The Lobby to have an ordering such as
\[
w_1 = w_2 = \cdots = w_n.
\]
If this is the case,
we see that we are left with an instance of $\plp{X}{Y}$, where
$\textsc{X} \in \{\textsc{MB},
\textsc{VB}\}$ and
$\textsc{Y} \in \{\textsc{SM},\textsc{AM}\}$.

We now introduce the
four probabilistic lobbying problems with issue weighting.
For
$\textsc{X} \in \{\textsc{MB},
\textsc{VB}\}$ and
$\textsc{Y} \in \{\textsc{SM},\textsc{AM}\}$, we define the
following problem.

\begin{description}
\item[Name:] \textsc{X-Y Probabilistic Lobbying Problem with
Issue Weighting}.
\item[Given:] A probability matrix $P \in
\matrixrationalszeroone{m}{n}$ with cost matrix $C_P$,
an issue weight vector
$\vec{W} \in \nats_{>0}^{n}$, an objective value $O \in \nats_{>0}$,
a budget~$B$, and some threshold $t\in \mathbb{Q}_{[0,1[}$.
\item[Question:] Is there a way for The Lobby to influence $C_P$ and
  hence $P$ (using bribery method
  \textsc{X} and evaluation criterion
\textsc{Y}, without exceeding budget $B$) such that the total
  weight of all issues for which the result coincides with The Lobby's
  target vector $1^n$ is at least~$O$?
\end{description}

We abbreviate this problem name as $\plpiw{X}{Y}$.

\section{Complexity-Theoretic Notions}
\label{sec:background-parameterized-complexity}

We assume the reader is familiar with standard notions of (classical)
complexity theory, such as $\p$, $\np$, and $\np$-completeness.  Since
we analyze the problems stated in
Section~\ref{sec:models-for-probabilistic-lobbying} not only in terms
of their classical complexity, but also with regard to their
\emph{parameterized} complexity,
we provide some basic notions here
(see, e.g., the text books by
Downey and Fellows~\cite{dow-fel:b:parameterized-complexity}, Flum and
Grohe~\cite{flu-gro:b:parameterized-complexity}, and
Niedermeier~\cite{nie:b:fixed-parameter-algorithms}
for more background).  As we derive our
results in a rather specific fashion, we will employ the ``Turing
way'' as proposed by
Cesati~\cite{ces:j:turing-way-parameterized-complexity}.

A \emph{parameterized problem} $\mathcal{P}$ is a subset of $\Sigma^*
{\times} \nats$, where $\Sigma$ is a fixed alphabet and $\Sigma^*$
is the set of strings over~$\Sigma$.
Each instance of the
parameterized problem $\mathcal{P}$ is a pair $(I, p)$, where the
second component $p$ is called the {\it parameter}. The language
$L(\mathcal{P})$ is the set of all \YES\ instances of~$\mathcal{P}$. 
 The parameterized problem $\mathcal{P}$ is {\it
fixed-parameter tractable} if there is an algorithm (realizable by a
deterministic Turing machine) that decides whether an input $(I, p)$
is a member of $L(\mathcal{P})$ in time $f(p)|I|^c$, where $c$ is a
fixed constant and $f$ is a function of the parameter~$p$, but
is independent
of the overall input length, $|I|$. The class of all fixed-parameter
tractable problems is denoted by $\FPT$.

The $\Oh^*(\cdot)$ notation has by now become standard in
 exact
algorithms.  It neglects not only constants (as the more
familiar $\Oh(\cdot)$ notation does) but also polynomial factors in the
function estimates.  Thus, a problem is in $\FPT$ if and only if an
instance (with parameter $p$) can be solved in time $\Oh^*(f(p))$ for
some %
function~$f$.

Sometimes, more than one parameter (e.g., two parameters $(p_1,p_2)$)
are associated with a (classical) problem.  This is formally
captured in the definition above by coding those parameters into one
number $p$ via a so-called pairing function through diagonalization.
As is standard, we assume our pairing function to be a polynomial-time
computable bijection from $\nats {\times} \nats$ onto $\nats$ that has
polynomial-time computable inverses.

For a given classical decision or optimization problem, there are
various ways to define parameters. %
With minimization problems, the \emph{standard parameterization} is a
bound on the entity to be minimized.  For instance, the problems
studied in this paper have, as a natural minimization objective, the
goal to minimize costs (i.e., to use a budget $B$ as small as
possible).
If one can assume that the parameter $p$ is small in practice, or in
practical situations involving humans, we can argue that an algorithm
offering a run time of $\Oh^*(2^p)$ behaves reasonably well in practice.
A related natural parameter
choice would be the budget that can be spent per issue, i.e., the
entity $\nicefrac{B}{n}$.  If the voters are actual human beings, one
can also argue that the discretization level $k$ would not be too
large.

One of the current trends in parameterized complexity analysis is to
study multiple parameterizations for each problem, including combining
multiple parameters for a problem instance.  This trend is highlighted
by two recent invited talks given by Fellows~\cite{Fel2009}
and Niedermeier~\cite{Nie2010}.
Notice that the study of different and multiple parameterizations can also be 
seen from another angle: 
Apart from identifying the hard parts of the problem instance, such research
represents a natural mathematical counterpart of the more
practically oriented quest for good parameters that may lead to the most
competitive algorithm frameworks for hard problems, as exemplified
most notably by SATzilla and ParamILS in the areas of algorithms for
Satisfiability and Integer Linear Programming, respectively, see
\cite{xu-hut-hoo-ley:j:SATzilla,hut-hoo-ley-stu:j:ParamILS}.

There is also a theory of parameterized complexity, as exhibited in 
\cite{dow-fel:b:parameterized-complexity,flu-gro:b:parameterized-complexity,nie:b:fixed-parameter-algorithms},
where parameterized complexity is expressed via hardness for or completeness
in the levels $\wt{t}$, $t \geq 1$, of the W-hierarchy,
which complement fixed-parameter tractability:
\[
\FPT = \wt{0} \subseteq \wt{1} \subseteq \wt{2} \subseteq \cdots.
\]
It is commonly believed that this hierarchy is strict.
Since only the second level, $\wtwo$, will be of interest to us
in this paper, we will define only this class below.

\begin{definition}
Let $\mathcal{P}$ and $\mathcal{P}'$ be two parameterized problems.
A \emph{parameterized reduction from $\mathcal{P}$ to $\mathcal{P}'$}
is a function $r$ that, for some polynomial $q$ and some function $g$, 
is computable in  time $\Oh(g(p)q(\problemsize{I}))$
 and maps an instance
$(I,p)$ of  $\mathcal{P}$ to an instance
$r(I,p)=(I',p')$ of  $\mathcal{P}'$
such that 
\begin{enumerate}
\item $(I,p)$ is a \YES\ instance of  $\mathcal{P}$ if and only if 
$(I',p')$  is a \YES\ instance of  $\mathcal{P}'$,
and 
\item $p'\leq g(p)$.
\end{enumerate}
We then say that $\mathcal{P}$ \emph{parameterized reduces
  to~$\mathcal{P}'$ (via~$r$)}.  Parameterized hardness for and completeness in
a parameterized complexity class is defined via parameterized
reductions.  We will show only $\wtwo$-completeness results.  A
parameterized problem $\mathcal{P}'$ is said to be \emph{$\wtwo$-hard}
if every parameterized problem $\mathcal{P}$ in $\wtwo$ parameterized
reduces to~$\mathcal{P}'$.  $\mathcal{P}'$ is said to be
\emph{$\wtwo$-complete} if $\mathcal{P}'$ is in $\wtwo$ and is
$\wtwo$-hard.
\end{definition}

$\wtwo$ can be characterized by the following problem on Turing machines:
\begin{description}
\item[Name:] \textsc{Short Multi-tape Nondeterministic Turing Machine
 Computation}.
\item[Given:] A multi-tape nondeterministic Turing machine $M$ (with
 two-way infinite tapes) and an input string $x$ (both $M$ and $x$ are
 given in some standard encoding).
\item[Parameter:] A positive integer $k$.
\item[Question:] Is there an accepting computation of $M$ on input $x$
 that reaches a final accepting state in at most $k$ steps?
\end{description}

We abbreviate this problem name as $\textsc{SMNTMC}$.

More specifically, a parameterized problem $\mathcal{P}$ is in $\wtwo$
if and only if it can be reduced to \textsc{SMNTMC} via a
parameterized reduction
\cite{ces:j:turing-way-parameterized-complexity}.  This can be
accomplished by giving an appropriate multi-tape nondeterministic
Turing machine for solving~$\mathcal{P}$.  Hardness for $\wtwo$
can be shown by
giving a parameterized reduction in the opposite direction, from
\textsc{SMNTMC} to~$\mathcal{P}$.

For
other applications of fixed-parameter tractability and parameterized
complexity to problems from computational social choice, see,
e.g.,~\cite{lin-rot:mpg:fpt-comsoc}.

\section{Classical Complexity Results}
\label{sec:classical-complexity}

We now provide a formal complexity analysis of the
probabilistic lobbying problems for all
combinations of bribery methods $\textsc{X} \in \{\textsc{MB},
\textsc{VB}\}$ and evaluation criteria
$\textsc{Y} \in \{\textsc{SM},\textsc{AM}\}$.
Table~\ref{tab:plp} summarizes some of our (classical and parameterized)
complexity results for
the problems $\plp{X}{Y}$; note that Table~\ref{tab:plp} does not
cover Theorem~\ref{thm:plpw2-b+k} (which considers a combination of
two parameters, namely of budget per issue and discretization level).

Some of these results are known from
previous work by
Christian et al.~\cite{chr-fel-ros-sli:j:lobbying},
as will be mentioned below.  Our results generalize
theirs by extending the model to probabilistic settings.
The listed $\FPT$ results might look peculiar at first glance, since
Christian et al.~\cite{chr-fel-ros-sli:j:lobbying} derived $\wtwo$-hardness
results, but this is due to the chosen parameterization,
as will be discussed later
in more detail.
We put parentheses around some classes in Table~\ref{tab:plp}
to indicate that these results
are trivially inherited from others. For example, if some problem is solvable
in polynomial time, then it is in $\FPT$ for any parameterization.
The table mainly provides results on the containment of problems in certain
complexity classes; if known, additional hardness results are also listed.

In Section~\ref{sec:classical-complexity:microbribery} we present our
results on microbribery (i.e., we study the problems
$\plp{\textsc{MB}}{Y}$ for $\textsc{Y} \in
\{\textsc{SM},\textsc{AM}\}$), and in Section~\ref{sec:voter-bribery}
we are concerned with voter bribery (i.e., we study the problems
$\plp{\textsc{VB}}{Y}$ for $\textsc{Y} \in
\{\textsc{SM},\textsc{AM}\}$).  In addition,
in Section~\ref{sec:plp-wiw} we study probabilistic lobbying with
issue weighting.

\begin{table}
\centering
\footnotesize
\begin{tabular}{|l||c|c|c|c||c|}
\hline
\multicolumn{1}{|c||}{Problem} & Classical  &
 \multicolumn{3}{c||}{Parameterized Complexity, parameterized by} &
 Stated in or 
 \\ \cline{3-5}
   & Complexity & Budget & Budget    & Budget \& & implied by 
 \\ 
   &            &        & per Issue & Discretiz.\ Level & Thm./Cor.
 \\\hline \hline
$\plp{\textsc{MB}}{\textsc{SM}}$
 & $\p$           & $(\FPT)$ & $(\FPT)$           & $(\FPT)$ &
\ref{thm:B1C1PLP} \\ \hline
$\plp{\textsc{MB}}{\textsc{AM}}$
 & $\p$           & $(\FPT)$ & $(\FPT)$           & $(\FPT)$ & 
\ref{thm:plp12-is-in-p} \\ \hline
$\plp{\textsc{VB}}{\textsc{SM}}$
 & $\np$-complete & $\FPT$ & $\wtwo$-complete & $\FPT$ & 
\ref{thm:plpnp}, \ref{thm:plp31-FPT}, \ref{thm:plpw2-a}, \ref{thm:plp3j-FPT}
\\ \hline
$\plp{\textsc{VB}}{\textsc{AM}}$
 & $\np$-complete & ? & $\wtwo$-hard & $\FPT$ & 
\ref{thm:plpnp}, \ref{cor:plpw2-a}, \ref{thm:plp3j-FPT} \\ \hline
\end{tabular}
\caption{Complexity results for $\plp{X}{Y}$, where
$\textsc{X} \in \{\textsc{MB},
\textsc{VB}\}$ and
$\textsc{Y} \in \{\textsc{SM},\textsc{AM}\}$}
\label{tab:plp}
\end{table}

\subsection{Microbribery}
\label{sec:classical-complexity:microbribery}

\begin{theorem}
\label{thm:B1C1PLP}
$\plp{MB}{SM}$ is in~$\p$.
\end{theorem}

\begin{proofs}
The aim is to win all referenda.
For each voter $v_i$  and referendum~$r_j$, $1 \leq i \leq m$ and
$1 \leq j \leq n$, we
compute in polynomial time the amount $b(v_i,r_j)$ The Lobby has to
spend to turn the favor of $v_i$ in the direction of The Lobby (beyond
the given threshold~$t$).  In particular, set $b(v_i,r_j)=0$ if voter $v_i$
would already vote according to the agenda of The Lobby.  For each
issue $r_j$, sort $\{b(v_i,r_j)\condition 1\leq i\leq m\}$ nondecreasingly, 
yielding a sequence $b_1(r_j)$, \dots, $b_m(r_j)$ such that
$b_k(r_j) \leq b_{\ell}(r_j)$ for $k < \ell$.
To win referendum~$r_j$, The Lobby must spend at least
$B(r_j)=\sum_{i=1}^{\lceil\nicefrac{(m+1)}{2}\rceil}b_i(r_j)$ dollars.
Hence, all referenda are won if and only if $\sum_{j=1}^n B(r_j)$ is
at most the given bribery budget~$B$.~\end{proofs}

Note that the time needed to implement the
algorithm given in the previous proof
can be bounded by a polynomial of low order.
More precisely, if the input consists of $m$ voters, $n$ referenda,
and discretization level~$k$,
then $\Oh(n\cdot m\cdot k)$ time is needed to compute the $b(v_i,r_j)$.
Having these values, $\Oh(n\cdot m\cdot \log m)$ time is needed for the sorting phase.
The sums can be computed in time $\Oh(n\cdot m)$.

Similarly, the other problems that
we show to belong to $\p$ admit solution algorithms
bounded by polynomials of low order.

The complexity of microbribery with evaluation criterion
\textsc{AM} is somewhat harder to determine.
We use the following auxiliary
problem.  First we need a definition.

\begin{definition}
\label{def:scheduling}
\begin{itemize}
\item  Given a directed graph $G$ consisting of path components
  $P_1,\ldots ,P_\pi$ with vertex set $V = \{J_1, \ldots , J_n\}$
  representing jobs,
  a \emph{schedule} $S$ of $q \leq n$ jobs (on a single
  machine) is a sequence $J_{i(1)}, \ldots , J_{i(q)}$ such that
  $J_{i(r)}=J_{i(s)}$ implies $r=s$ (i.e., no job appears twice).

\item
  Assigning cost $c(J_k)$ to job $J_k$ for each~$k$, $1 \leq k \leq n$,
  the \emph{cost of schedule $S$} is
\[
c(S)=\sum_{k=1}^q c(J_{i(k)}).
\]

\item The mapping $\phi$ we use in the following takes two arguments:
the index of a path and the number of the job on that path, and 
it returns the correct index of the job.
  $S$ is said to \emph{respect the
  precedence constraints of $G$} if for every path-component
  $P_i=(J_{\phi(i,1)}, \ldots , J_{\phi(i,p(i))})$ of $G$ (with
  $V = \bigcup_{i=1}^\pi\{J_{\phi(i,\ell)} \mid 1 \le \ell \le p(i)\}$)
  and for each $\ell$ with $2 \le \ell
  \le p(i)$, we have: If $J_{\phi(i,\ell)}$ occurs in schedule $S$ then
  $J_{\phi(i,\ell-1)}$ occurs in $S$ before $J_{\phi(i,\ell)}$.
\end{itemize}
\end{definition}

\begin{description}
\item[Name:] {\sc Path Schedule}
\item[Given:] A set $V=\{J_1, \ldots , J_n\}$ of jobs, a directed
  graph $G=(V,A)$ consisting of pairwise disjoint paths $P_1,\ldots,
  P_\pi$, two numbers $C,q \in \nats$, and
a cost function $c:V \to \nats$.
\item[Question:]
  Does there exist a schedule $J_{i(1)},\ldots,J_{i(q)}$ of $q$ jobs
  of cost at most $C$ respecting the precedence constraints of~$G$?
\end{description}

We first show, as Lemma~\ref{lem:mps}, that the minimization version of
{\sc Path Schedule} is polynomial-time computable, so the decision
problem {\sc Path Schedule} is in~$\p$.
Then we will show, as Theorem~\ref{thm:plp12-is-in-p}, how to
reduce $\plp{MB}{AM}$ to {\sc Path Schedule}, which implies that
$\plp{MB}{AM}$ is in $\p$ as well.

\begin{lemma}
\label{lem:mps}
\textsc{Path Schedule} is in~$\p$.
\end{lemma}

\begin{proofs}
Given an instance of \textsc{Path Schedule} as in the problem
description above, the following dynamic programming approach calculates
$T[\{P_1,\ldots ,P_\pi\},q]$, which gives the minimum cost to solve the
problem.
 
We build up a table $T[\{P_1,\ldots ,P_\ell\},j]$ storing the
minimum cost of scheduling $j$ jobs out of the jobs contained in the
paths $P_1,\ldots, P_\ell$.
Let 
$P_i=(J_{\phi(i,1)}, \ldots , J_{\phi(i,p(i))})$
be a path, $1 \leq i \leq \pi$. 
For $k\leq p(1)$, set
$T[\{P_1\},k]=\sum_{s=1}^kc(J_{\phi(1,s)})$. 
For $k>p(1)$, set $T[\{P_1\},k]=\infty$. 
If $\ell>1$, 
$T[\{P_1,\ldots,P_\ell\},j]$ equals
\[
\min_{0 \le k \le \min\{j,p(\ell)\}} T[\{P_1,\ldots ,P_{\ell-1}\},j-k]
+ \sum_{s=1}^k c(J_{\phi(\ell,s)}).
\]
Consider each possible scheduling of the first
$k$ jobs of~$P_\ell$.  For the remaining $j-k$ jobs, look up a
solution in the table.  
Note that we can re-order each schedule $S$ so that all jobs from
one path contiguously appear in $S$, without violating the precedence
constraints by this re-ordering nor changing the cost of the schedule. 
Hence, $T[\{P_1,\ldots ,P_\pi\},q]$ gives the minimum schedule cost. 
The number of entries in the table is $\pi\cdot q$, and computing each
entry $T[\{P_1,\dots, P_\ell\},\cdot]$ is linear in $p(\ell)$ (for
each~$\ell$, $1\leq \ell\leq \pi$), which leads to a run time of the
dynamic programming algorithm that is polynomially bounded in the
input size.~\end{proofs}

\begin{theorem}
\label{thm:plp12-is-in-p}
$\plp{MB}{AM}$ is in~$\p$.
\end{theorem}

\begin{proofs}
Let $(P,C_P,
B,t)$ be a given $\plp{MB}{AM}$ instance, where $P \in
\matrixrationalszeroone{m}{n}$, $C_P$ is a
cost matrix,
$B$ is The Lobby's budget, and $t$ is a given threshold.
Let $k$ be the discretization level of~$P$, i.e., the interval
is divided into $k+1$ steps of size $\nicefrac{1}{(k+1)}$ each.
For $j \in \{1, 2, \ldots, n\}$, let $d_j$ be the minimum
cost for The Lobby to bring referendum $r_j$ into line with the $j$th
entry of its target vector~$1^n$.  If $\sum_{j=1}^n d_j \le B$,
then The Lobby can achieve its goal that the votes on
all issues
pass.
For every~$r_j$, create an
equivalent \textsc{Path Scheduling} instance.  First, compute for
$r_j$ the minimum number $b_j$ of bribery steps needed to achieve The
Lobby's goal on~$r_j$.  That is, choose the smallest $b_j \in
\nats$ such that $\overline{p_j}+\nicefrac{b_j}{(k+1)m} >t$.
Now, given~$r_j$, derive a path $P_i$ from the price
function $c_{i,j}$ for every voter~$v_i$, $1 \leq i \leq m$, as follows.
\begin{enumerate}
\item Let $s$, $0 \le s \le k+1$, be minimum with the property
  $c_{i,j}(\nicefrac{s}{(k+1)})\in \nats_{>0}$.

\item Create a path $P_i = ((p_s,i), \ldots , (p_{k+1},i))$, where
$p_h = \nicefrac{h}{(k+1)}$.

\item Assign the cost $\hat c((p_h,i))=
  c_{i,j}(p_h) - c_{i,j}(p_{(h-1)})$ to~$(p_h,i)$, where $s+1 \leq h \leq k+1$.
\end{enumerate}

Note that $\hat c((p_h,i))$
represents the cost of raising the probability of voting ``yes'' from
$\nicefrac{(h-1)}{(k+1)}$ to $\nicefrac{h}{(k+1)}$.  In order to do
so, we must have reached
an acceptance probability of $\nicefrac{(h-1)}{(k+1)}$ first.  Now,
let the number of jobs to be scheduled be~$b_j$.
Note that one can take $b_j$ bribery steps at
the cost of $d_j$ dollars if and only if one can schedule $b_j$ jobs
with a cost of~$d_j$. Hence, we can decide whether or not
$(P,C_P,
B)$ is in $\plp{MB}{AM}$ by using the dynamic program
given in the proof of Lemma~\ref{lem:mps}.~\end{proofs}

\subsection{Voter Bribery}
\label{sec:voter-bribery}

Recall the \textsc{Optimal Lobbying} problem ($\ol$) defined in
Section~\ref{sec:introduction-motivation}.
Again, The Lobby's target
vector $\vec{Z}$ may assumed to be all ones, without loss of
generality, so $\vec{Z}$ may be dropped from the input.

Christian et al.~\cite{chr-fel-ros-sli:j:lobbying} proved that this problem
is $\wtwo$-complete by reducing from the $\wtwo$-complete
problem $k$-\textsc{Dominating Set}
to $\ol$ (showing the lower bound) and from $\ol$ to
the $\wtwo$-complete problem 
\textsc{Independent}-$k$-\textsc{Dominating Set} (showing the upper bound).
In particular, this implies $\np$-hardness of $\ol$.

The following
result focuses on the classical complexity of $\plp{VB}{SM}$ and
$\plp{VB}{AM}$; the parameterized complexity of these problems will be
studied in Section~\ref{sec:parameterized-complexity} and will make
use of the proof of Theorem~\ref{thm:plpnp} below.

To employ
the $\wtwo$-hardness result of Christian et al.~\cite{chr-fel-ros-sli:j:lobbying}, 
we show that $\ol$ is a special case of
$\plp{VB}{SM}$ and thus (parameterized) polynomial-time reduces
to $\plp{VB}{SM}$.
Analogous arguments apply to $\plp{VB}{AM}$.

\begin{theorem}
$\plp{VB}{SM}$ and $\plp{VB}{AM}$ are $\np$-complete.
\label{thm:plpnp}
\end{theorem}

\begin{proofs}
Membership in $\np$ is obtained through a ``guess-and-check'' 
algorithm for $\plp{VB}{SM}$ and
$\plp{VB}{AM}$.  We present the details for the sake of completeness.  Let
$(P,C_P,B,t)$ be a given instance of $\plp{VB}{Y}$ for evaluation
criterion $\textsc{Y} \in \{\textsc{SM},\textsc{AM}\}$, where $P \in
\matrixrationalszeroone{m}{n}$ is a probability matrix with cost
matrix $C_P$ (with integer entries), $B$ is a budget, and $t$ is a
threshold.
Nondeterministically guess a subset $V$ of rows in $P$ (each
corresponding to a voter to be influenced in this nondeterministic
branch) and a corresponding list $D_V = (d_1, d_2, \dots , d_{\|V\|})$
of integers such that $\sum_{i=1}^{\|V\|} d_i \leq B$, where $d_i$ is
the amount of bribery money to be spent on $v_i \in V$ in this
nondeterministic branch.  For any $(V,D_V)$ guessed, check
deterministically whether spending  $d_i$ on $v_i$, for
each $v_i \in V$, will change (according to the cost matrix $C_P$) the
given matrix $P$ into a new matrix $P_V$ such that each issue
evaluates to ``yes'' in $P_V$ under evaluation criterion $\textsc{Y}$
with respect to threshold~$t$,
and accept/reject accordingly on this nondeterministic branch.

We
now prove that $\plp{VB}{SM}$ is $\np$-hard
by reducing
$\ol$ to $\plp{VB}{SM}$.
We are given an instance
$(E,b)$ of $\ol$, where $E$ is a $m
{\times} n$ $0/1$ matrix and $b$ is the number of votes to be
edited.  Recall that The Lobby's target vector is~$1^n$.
We construct an instance of $\plp{VB}{SM}$ consisting of the given
matrix $P = E$ (a ``degenerate'' probability matrix with only the
probabilities $0$ and~$1$), a corresponding cost matrix~$C_P$,
and a budget~$B$.  $C_P$ has two
columns (we have $k=0$, since the problem instance is
deterministic, see Section~\ref{sec:initialmodel}), one column for
probability $0$ and one for probability~$1$.  All entries of $C_P$ 
corresponding to $p_{i,j}\neq 1$ are set to unit cost:  
$c_{i,j}[1]=1$ if $p_{i,j}\neq 1$.  Set the threshold $t$ 
to~$\nicefrac{1}{2}$.

The cost of increasing any value in $P$ is $n$, since donations are
distributed evenly across issues for a given voter.  We want to know
whether there is a set of bribes of cost at most $b\cdot n=B$ such that
The Lobby's agenda passes.  This holds if and only if there are $b$
voters that can be bribed so that they vote uniformly according to The
Lobby's agenda and that is sufficient to pass all the issues.  Thus,
the given instance $(E,b)$ is in $\ol$ if and only if the
constructed instance $(P, C_P,
B,t)$ is in $\plp{VB}{SM}$, which shows that $\ol$ is a polynomial-time
recognizable special case of $\plp{VB}{SM}$, and thus $\plp{VB}{SM}$
is $\np$-hard.

Note that for the construction above it does not matter whether we use
the strict-majority criterion
(\textsc{SM}) or the average-majority
criterion
(\textsc{AM}).  Since the entries of $P$ are $0$ or~$1$,
we have $\overline{p_{j}} > 0.5$ if and only if we have a strict
majority of ones in the $j$th column.  Thus, $\plp{VB}{AM}$ is
$\np$-hard, too.~\end{proofs}

\subsection{Probabilistic Lobbying with Issue Weighting}
\label{sec:plp-wiw}

\begin{table}
\centering
\footnotesize
\begin{tabular}{|l||c|c|c|c||c|}
\hline
\multicolumn{1}{|c||}{Problem} & Classical 
 & \multicolumn{3}{c||}{Parameterized Complexity}
 &
 Stated in or 
 \\ \cline{3-5}
   & Complexity & Budget & Budget    & Budget \& & implied by 
 \\ 
   &            &        & per Issue & Discr.\ Level & Thm./Cor.
 \\\hline \hline
$\plpiw{MB}{SM}$ & $\np$-complete & $\FPT$ & ?& $(\FPT)$ & 
\ref{thm:plpiw-11-12-21-22-is-np-hard}, \ref{thm-issueweighted-FPT} \\ \hline
$\plpiw{MB}{AM}$ & $\np$-complete & $\FPT$ & ?& $(\FPT)$ & 
\ref{thm:plpiw-11-12-21-22-is-np-hard}, \ref{thm-issueweighted-FPT} \\ \hline
$\plpiw{VB}{SM}$ & $\np$-complete & $\FPT$ & $\wtwo$-complete$^*$ & $\FPT$ & 
\ref{cor:plpnp},
\ref{cor:vb-sm-plp-wiw-vb-am-plp-wiw-budget-and-disc-level},
\ref{cor:wiw-wtwo-hard}, \ref{thm:plpiw31}
 \\ \hline
$\plpiw{VB}{AM}$ & $\np$-complete & ? & $\wtwo$-hard & $\FPT$ & 
\ref{cor:plpnp},
\ref{cor:wiw-wtwo-hard},
\ref{cor:vb-sm-plp-wiw-vb-am-plp-wiw-budget-and-disc-level}
 \\ \hline
\end{tabular}
\caption{Complexity results for $\plpiw{X}{Y}$, where
$\textsc{X} \in \{\textsc{MB},
\textsc{VB}\}$ and
$\textsc{Y} \in \{\textsc{SM},\textsc{AM}\}$}
\label{tab:plpiw}
\end{table}

Table~\ref{tab:plpiw} summarizes some of our results for
$\plpiw{X}{Y}$, where
$\textsc{X} \in \{\textsc{MB},
\textsc{VB}\}$ and
$\textsc{Y} \in \{\textsc{SM},\textsc{AM}\}$;
again, note that Table~\ref{tab:plpiw} does not
cover all our results.  The most interesting
observation from the table
is that introducing issue weights raises the complexity from
$\p$ to $\np$-completeness for all cases of microbribery
(though it remains the same for voter bribery).  Nonetheless,
we show  (Theorem~\ref{thm-issueweighted-FPT})
that these $\np$-complete problems are fixed-parameter tractable.
Another interesting observation concerns the question of membership in $\wtwo$.
In the case indicated by the $*$ annotation, we can show this membership 
only when we take the lower bound $O$ quantifying the objective of the bribery
(in terms of issue weights) as a further parameter.
Question marks indicate open problems.

\begin{theorem}
\label{thm:plpiw-11-12-21-22-is-np-hard}
$\plpiw{MB}{SM}$ and  $\plpiw{MB}{AM}$
are each $\np$-com\-plete.
\end{theorem}

\begin{proofs}
Membership in $\np$ can be shown with a ``guess-and-check'' 
algorithm for both problems..
The argument is analogous to that for $\np$ membership of the problems
$\plpiw{VB}{SM}$ and $\plpiw{VB}{AM}$ presented in the proof of
Theorem~\ref{thm:plpnp}, except that now we guess a subset $E$ of
entries in the given probability matrix $P$ (rather than a subset $V$
of rows of~$P$) along with a corresponding list $D_E = (d_1, d_2,
\dots , d_{\|E\|})$ that collects the amounts of bribery money to be
spent on each entry in~$E$.

To prove that $\plpiw{MB}{SM}$ is $\np$-hard, we give a reduction from the
well-known $\np$-complete problem
\textsc{Knapsack} (see, e.g., \cite{gar-joh:b:int}) to the problem
$\plpiw{MB}{SM}$.
In \textsc{Knapsack}, we are
given a set of objects $U=\{o_1,\ldots , o_n\}$ with weights $w:U \to
\nats$ and profits $p:U \to \nats$, and $W,P \in
\nats$.  The question is whether there is a subset $I \subseteq
\{1,\ldots ,n\}$ such that $\sum_{i \in I} w(o_i) \le W$ and $\sum_{i
  \in I} p(o_i) \ge P$.  Given a \textsc{Knapsack} instance
$(U,w,p,W,P)$, create an $\plpiw{MB}{SM}$ instance with $k=0$ and only
one voter, $v_1$,
where for each issue, $v_1$'s acceptance probability
is either zero or one.  For each object~$o_j \in U$, create
an issue $r_j$ such that the acceptance probability of $v_1$ is
zero. Let the cost of raising this probability on $r_j$ be
$c_{1,j}(1)=w(o_j)$ and let the weight of issue $r_j$ be $w_j=p(o_j)$.  Let
The Lobby's budget be $W$ and its objective value be $O=P$.
Set the threshold $t$ to~$\nicefrac{1}{2}$.  By construction,
there is a subset $I \subseteq \{1,\ldots ,n\}$ with $\sum_{i \in I}
w(o_i) \le W$ and $\sum_{i \in I} p(o_i) \ge P$ if and only if there
is a subset $I \subseteq \{1,\ldots ,n\}$ with $\sum_{i \in I} c_{1,i}(1) \le
W$ and $\sum_{i \in I} w_i \ge O$.

As the reduction introduces only one voter, there is no difference
between the evaluation criteria
\textsc{SM} and
\textsc{AM}.  Hence, the above reduction works for 
both problems.~\end{proofs}

Turning now to voter bribery with issue weighting, note that an
immediate consequence of Theorem~\ref{thm:plpnp} is that
$\plpiw{VB}{SM}$ and $\plpiw{VB}{AM}$ are
$\np$-hard, since they are generalizations of $\plp{VB}{SM}$ and
$\plp{VB}{AM}$, respectively.  Again, membership in
$\np$ can be seen using appropriate ``guess-and-check'' algorithms for the more
general problems.

\begin{corollary}
  $\plpiw{VB}{SM}$ and $\plpiw{VB}{AM}$ each are
  $\np$-com\-plete.
\label{cor:plpnp}
\end{corollary}

\section{Parameterized Complexity Results}
\label{sec:parameterized-complexity}

In this section, we study the parameterized complexity of our
probabilistic lobbying problems.  Parameterized hardness is usually
shown by proving hardness for the levels of the W-hierarchy (with
respect to parameterized reductions).  Indeed, this hierarchy may be
viewed as a ``barometer of parametric intractability''
\cite[p.~14]{dow-fel:b:parameterized-complexity}.  The lowest two
levels of the W-hierarchy, $\wt{0} = \FPT$ and $\wt{1}$, are the
parameterized analogues of the classical complexity classes $\p$
and~$\np$.  We will show completeness results for the $\wtwo$ level of
this hierarchy.

In parameterized complexity, the standard parameterization for minimization
problems is an upper bound on the entity to be minimized. In our case,  
this is the budget~$B$. Since in the voter bribery model, the money is  
equally distributed over all referenda, it also makes sense to consider the  
upper bound $\nicefrac{B}{n}$, i.e., the budget per referendum, 
as a natural, derived parameter.
Another natural way of parameterization is derived from certain  
properties of the input, be they implicit or explicit.  In our case,  
the discretization level can be considered as such a parameter, in  
particular, since the smallest discretization level has been already  
considered before within the \textsc{optimal lobbying}
problem~\cite{chr-fel-ros-sli:j:lobbying}.
Therefore, we examine all three of these parameterizations in order to
understand the effect the choice of parameterizations has on the
complexity of the problems.

\subsection{Voter Bribery}

\begin{theorem}
$\plp{VB}{SM}$ and $\plp{VB}{AM}$
(parameterized by the budget and by the discretization level) are in 
$\FPT$.
\label{thm:plp3j-FPT}
\end{theorem}

\begin{proofs}
Consider an instance of  $\plp{VB}{Y}$,
$\textsc{Y} \in \{\textsc{SM},\textsc{AM}\}$,
i.e., we are given $n$ referenda and
$m$ voters, as well as a cost matrix $C_P$ (with either
$--$ or integer entries),
a discretization level~$k$, a budget~$B$, and a threshold~$t$.
Recall that the target vector $\vec{Z}$ of The Lobby is assumed to
be~$1^n$.  Hence, the rows of $C_P$ are monotonically
nondecreasing (after possibly some $--$ entries). 
Observe that any successful
bribe of any voter needs at least $n$ dollars, since 
the money is evenly distributed
among all referenda, and at least one dollar is needed to
influence the chosen voter's votes for all referenda. Hence, 
$B\geq n$.
We can assume that any entry in $C_P$ is limited by $B+1$, 
after replacing every entry bigger than $B$ by $B+1$.
Notice that the entry $B+1$ reflects that the
intended bribery cannot be afforded.

Although $k$ could be bigger than $B$, the interesting
area of each row in $C_P$ (containing integer entries) 
cannot have more than $B$ strict increases in the sequence.
We therefore encode each row in $C_P$ by a sequence
$(k_1, b_1, k_2, b_2, \dots, k_\ell, b_\ell)$,
$\ell\leq B$, which reads as follows:
By investing $b_j$  dollars, we proceed to column number 
$\sum_{i\leq j}k_i$.
Note that $k$ is given in unary in the original instance (implicitly
by giving the cost matrix $C_P$), and that each $k_j$ can be encoded
with $\log k$ bits.
Hence, we extract from $C_P$ for each voter $v$ a submatrix $S_P(v)$
with $n\leq B$ rows (for the referenda) and at most $2B$ columns
(encoding the ``jumps'' in the integer sequence as described above).
This matrix with at most $B$ rows and at most $2B$ columns can be 
alternatively viewed as a matrix with at most $B$ rows and at most $B$ 
columns, where 
each matrix entry consists of a pair of numbers,             
one between $1$ and $B+1$ and one of size at most $\log k$.
Therefore,
we can associate with
each voter at most $(B+1+ \log k)^{B^2}$ distinct submatrices $S_P(v)$ of
this kind, called voter profiles.
It makes no sense to store more than $B$ voters with the same profile.
Hence, we can assume that $m\leq B\cdot (B+1+\log k)^{B^2}$.
Therefore, all relevant parts of the input are bounded by a 
function in the parameters
$B$ and $k$,\footnote{In technical terms,
this means that we have derived a so-called problem kernel for this problem.}
so that a brute-force algorithm can be used to solve the instance.
This shows that the problem is in $\FPT$.~\end{proofs}
If we assume that the discretization level
is a rather small number, the preceding theorem
says that the problems $\plp{VB}{SM}$ and $\plp{VB}{AM}$
can be solved efficiently in practice.
Although we were not able to establish an
$\FPT$ result for $\plp{VB}{AM}$
when the discretization level
is not part of the parameter (but only the budget is),
we can overcome this formal obstacle for
$\plp{VB}{SM}$, as the following result shows.

\begin{theorem}
  $\plp{VB}{SM}$ (parameterized by the budget) is in $\FPT$.
\label{thm:plp31-FPT}
\end{theorem}

\begin{proofs}
Let an instance of $\plp{VB}{SM}$ be given.
From the given cost matrix $C_P$, we  extract
the information $W(i,j)$ that gives the minimum amount
of money The Lobby  must spend on voter $v_i$ to
turn his or her voting behavior on issue $r_j$ 
in favor of The Lobby's agenda,
eventually raising the corresponding voting probability beyond the
given threshold $t$. 
Each entry in $W(i,j)$ is between $0$ and $B$.
Moreover, as argued in the previous proof, there are no more than $B$ issues and we  again define a voter
profile (this time the $i$th row of the table $W(i,j)$ gives such a profile) 
for each voter, and we need to keep at most $B$ voters with the same profile.
Hence, no more than
$B(B+1)^{B}$ voters are present in the instance.
Therefore, some brute-force approach can be used to show membership in
$\FPT$.~\end{proofs}

The area of parameterized complexity leaves some freedom regarding the
choice of parameterization.  The main reason that the standard
parameterization (referring to the entity to be minimized, in this
case the budget) yields an $\FPT$ result is the fact that the
parameter is already very big compared to the overall input (e.g., the
number of issues $n$) by the very definition of the problem: Since the
money given to one voter will be evenly distributed among the issues
and since the cost matrix contains only integer entries, it makes no
sense at all to spend less than $n$ dollars on a voter.  Hence, the
budget should be at least $n$ dollars (assuming that some of the
voters must be influenced by The Lobby to achieve their agenda).  This
obstacle can be sidestepped
by changing the parameterization to $\nicefrac{B}{n}$, i.e., to
  the ``budget per issue'' (see, e.g., Theorem~\ref{thm:plpw2-a}).
Note that another way would be allowing rational numbers as entries in 
the cost matrix but we will not consider this in this paper but rather 
focus on the previous one.

\begin{theorem}
  $\plp{VB}{SM}$ %
(parameterized by the budget per
  issue) is $\wtwo$-complete.
\label{thm:plpw2-a}
\end{theorem}

\begin{proofs}
$\wtwo$-hardness can be derived from the proof of
Theorem~\ref{thm:plpnp}.  Recall that in the proof of
this theorem an instance $(E,b)$ of 
$\ol$ was reduced to an instance of $\plp{VB}{SM}$, 
with budget $B=n\cdot b$. Hence, the parameter ``budget per issue''
of that $\plp{VB}{SM}$ instance equals
$b$. Therefore, the reduction in  the proof of Theorem~\ref{thm:plpnp}
preserves the parameter and hence
$\wtwo$-hardness follows from the $\wtwo$-hardness of~$\ol$,
see~\cite{chr-fel-ros-sli:j:lobbying}.
Moreover, the instance of $\plp{VB}{SM}$ produced by the reduction has
discretization level zero.

To show membership in $\wtwo$, we reduce
$\plp{VB}{SM}$ to \textsc{SMNTMC}, which was defined in
Section~\ref{sec:background-parameterized-complexity}.  To this end,
it suffices to describe how a nondeterministic multi-tape Turing
machine can solve such a lobbying problem.

Consider an instance of $\plp{VB}{AM}$: a probability matrix $P \in
\matrixrationalszeroone{m}{n}$ with a
cost matrix~$C_P$,
a budget~$B$, and
a fixed threshold~$t$.  We may identify $t$ with a
certain step level for the
price functions.

The reducing machine works as follows.  From $P$, $C_P$, and $t$, the
machine extracts the information $H_{i,j}(d)$, $1\leq d \leq B$,
where $H_{i,j}(d)$ is
true if  $p_{i,j}\geq t$ or if $c_{i,j}(t)\leq \nicefrac{d}{n}$. 
Note that the bribery money is evenly distributed across all issues, also
note that $H_{i,j}(d)$ captures whether paying $d$ dollars
to voter $v_i$ helps to raise the acceptance probability of $v_i$ on
referendum $r_j$ above the threshold~$t$.  Moreover, for each
referendum~$r_j$, the reducing machine computes
the minimum number of voters that need to switch their opinion so that
majority is reached for that specific referendum; let $s(j)$ denote
this threshold for~$r_j$.  Since the cost matrix contains integer entries,
meaningfully bribing $s$ voters costs at least $s\cdot n$ dollars; only then
each referendum will receive at least one dollar per voter. 
Hence, a referendum with $s(j) > \nicefrac{B}{n}$ yields a
\NO\ instance. We can therefore replace any value $s(j) >
\nicefrac{B}{n}$ by the value $\lfloor \nicefrac{B}{n} \rfloor +1$.

From $H_{i,j}(d)$, the reducing machine produces (basically by sorting)
another \emph{winning table} $W_i(\ell)$ that lists for voter $v_i$ those referenda where 
the acceptance probability of $v_i$ on
referendum $r_j$ is raised above the threshold~$t$ by paying to $v_i$ the amount of $\ell\cdot n$ dollars but not
by paying $(\ell-1)\cdot n$ dollars.
Note that we can assume that the bribery money is spent in
multiples of $n$, the number of referenda, since spending $n$ dollars
on some voter means spending one dollar per issue for that voter.
This table is initialized by $W_i(0)$ listing those referenda already
won at the very beginning, although this is not an important issue due to the information contained in $s(j)$.

The nondeterministic multi-tape Turing machine $M$ we describe next
has, in particular, access to $W_i(\ell)$ 
 and to $s(j)$.  $M$ has $n+1$
working tapes $T_j$, $0\leq j\leq n$, all except one of which
correspond to issues~$r_j$, $1 \leq j \leq n$.  We will use the set of
voters, $V=\{v_1,\dots,v_m\}$, as part of the work alphabet.  The (formal) input tape
of $M$ is ignored.

$M$ starts by writing $s(j)$ symbols $\#$ onto tape~$j$ for each~$j$,
$1 \leq j \leq n$.  By using parallel writing steps, this needs at
most $\lfloor \nicefrac{B}{n} \rfloor+1$ steps, since $s(j)\leq\lfloor
\nicefrac{B}{n} \rfloor+1$ as argued above.
   We also need an ``information hiding'' trick here: every time the machine writes a $\#$ symbol, it moves
the writing head, so that in the next step the head will read a blank symbol. The trick is required in
order to keep the transition table small: basically, we cannot insert in the transition table $2^n$
different instructions to take into account all different conﬁgurations of blank and $\#$ symbols on the
$n$ tapes.

Second, for each $i \in \{1,\dots,m\}$, $M$ writes $k_i$ symbols $v_i$
from the alphabet $V$ on the zeroth tape, $T_0$, such that
$\sum_{i=1}^{m} k_i\leq \nicefrac{B}{n}$.  This is the
nondeterministic guessing phase where the amount of bribery money
spent on each voter, namely $k_i\cdot n$ for voter $v_i$, is
determined. The finite control is used to ensure that a word
from the language $\{v_1\}^*\cdot\{v_2\}^*\cdots\{v_m\}^*$ is written on tape~$T_0$. 

In the third phase, %
$M$
reads tape~$T_0$.
In its finite control, $M$ stores the ``current voter'' whose bribery money is read.
For each voter $v_i$, a counter $c_i$ is provided (within the finite memory of $M$).
If a symbol $v_i$ is read, $c_i$ is incremented,
and then  $M$ moves in parallel all the heads on the tapes~$T_j$,
where $j$ is contained in $W_i(c_i)$. 
Hence, the string on tape~$T_0$ is being processed in
at most $\nicefrac{B}{n}$ (parallel) steps.

Finally, it is checked if the left border is reached (again) for all
tapes~$T_j$, $j>0$.  This is the case if and only if the guessed
bribery was successful.~\end{proofs}

The $\wtwo$-hardness 
proof for $\plp{VB}{AM}$ 
is analogous.

Recall that
$\plp{VB}{SM}$
is the same as $\plp{VB}{AM}$ 
if the discretization level is zero. So, we conclude:

\begin{corollary}
$\plp{VB}{AM}$ 
(parameterized by the budget per
  issue) is $\wtwo$-hard.
\label{cor:plpw2-a}
\end{corollary}

Membership in $\wtwo$ is an open problem for $\plp{VB}{AM}$ 
when parameterized by the budget per issue.  In contrast, we show 
definitive parameterized complexity results for different
parameterizations.

\begin{theorem}
  $\plp{VB}{SM}$ and 
$\plp{VB}{AM}$ 
(parameterized by the budget per
  issue and by the discretization level) are  $\wtwo$-complete.
\label{thm:plpw2-b+k}
\end{theorem}

\begin{proofs}
 As mentioned above, we already have hardness for a discretization level of zero,
 i.e., when the second parameter is fixed to the lowest possible value.
We implicitly show membership in $\wtwo$ for 
  $\plp{VB}{SM}$ 
in Theorem~\ref{thm:plpw2-a}  when the second parameter only plays
a role in the polynomial part of the run time estimate. Hence, the claim is a simple
corollary of what we have already shown above.

It remains to prove membership in $\wtwo$ for 
  $\plp{VB}{AM}$.
This can be seen by modifying the proof of Theorem~\ref{thm:plpw2-a}. 
To show membership in $\wtwo$, we reduce
$\plp{VB}{AM}$ to \textsc{SMNTMC}.
So, 
it suffices to describe how a nondeterministic multi-tape Turing
machine can solve such a lobbying problem.

Consider an instance of $\plp{VB}{SM}$: a probability matrix $P \in
\matrixrationalszeroone{m}{n}$ with a
cost matrix~$C_P$,
a budget~$B$, and
a fixed threshold~$t$.  Again, we may identify $t$ with a
certain step level for the
price functions.

From this input, the reducing machine computes the following:

\begin{itemize}
\item It does some preprocessing, so that it is guaranteed that the overall money that could
be meaningfully invested on any voter is bounded by $\left\lceil\nicefrac{B}{n}\right\rceil$, which is  the first parameter.
No larger amount of money is available, anyhow.
Confidence steps that cannot be reached at all are modeled by requiring
an investment of (all in all) $\left\lfloor\nicefrac{B}{n}\right\rfloor+1$
dollars on that voter for each issue.
\item It computes the entity $s(j)$ that now denotes the number of confidence steps 	
issue $r_j$ has to be raised in total to ensure a win of that referendum.
Notice that $s(j)$ can be assumed to be bounded by the product of the first parameter, more precisely, by $\left\lfloor \nicefrac{B}{n}\right\rfloor+1$ (as
argued before), and the second parameter, more precisely, by $k+1$:
Each bribe (to whatever voter $v$) of $n$ dollars (recall that we can again rely on bribery money being
used in multiples of $n$, the number of issues) will
raise $v$'s confidence in voting
according to the agenda of The Lobby by at most $k+1$ steps.  

\item It finally computes
$W_i(\ell)$ which now gives the list of issues whose confidence is raised when investing $\ell\cdot n$ dollars
on $v_i$ (compared to $(\ell-1)\cdot n$ dollars), plus the number of confidence steps by which the corresponding
issue is raised. 
\end{itemize}

The nondeterministic multi-tape Turing machine $M$ we describe next
has, in particular, access to $W_i(\ell)$ 
 and to $s(j)$.  $M$ has $n+1$
working tapes $T_j$, $0\leq j\leq n$, all except one of which
correspond to issues~$r_j$, $1 \leq j \leq n$.  We will use the set of
voters, $V=\{v_1,\dots,v_m\}$, as part of the work alphabet.  The (formal) input tape
of $M$ is ignored.

$M$ starts by writing $s(j)$ symbols $\#$ onto tape~$j$ for each~$j$,
$1 \leq j \leq n$.  By using parallel writing steps, this needs at
most $f(\nicefrac{B}{n},k):=\left(\lfloor \nicefrac{B}{n} \rfloor+1\right)\left(k+1\right)$ steps, since $s(j)\leq f(\nicefrac{B}{n},k)$ as argued above.
The ``information hiding'' trick works as before.

Second, for each $i \in \{1,\dots,m\}$, $M$ writes $k_i$ symbols $v_i$
from the alphabet $V$ on the zeroth tape, $T_0$, such that
$\sum_{i=1}^{m} k_i\leq \nicefrac{B}{n}$.  This is the
nondeterministic guessing phase where the amount of bribery money
spent on each voter, namely $k_i\cdot n$ for voter $v_i$, is
determined. The finite control is used to ensure that a word
from the language $\{v_1\}^*\cdot\{v_2\}^*\cdots\{v_m\}^*$ is written on tape~$T_0$. 

In the third phase, %
$M$
reads tape~$T_0$.
In its finite control, $M$ stores the ``current voter'' whose bribery money is read.
For each voter $v_i$, a counter $c_i$ is provided (within the finite memory of $M$).
If a symbol $v_i$ is read, $c_i$ is incremented,
and then  $M$ moves in parallel all the heads on the tapes~$T_j$,
where $j$ is contained in $W_i(c_i)$; notice that the number of
steps each head has to move is now also stored in $W_i(c_i)$. 
Hence, the string on tape~$T_0$ is being processed in
at most $f\left(\nicefrac{B}{n},k\right)$ (parallel) steps.

Finally, it is checked if the left border is reached (again) for all
tapes~$T_j$, $j>0$.  This is the case if and only if the guessed
bribery was successful.~\end{proofs}

\subsection{Probabilistic Lobbying with Issue Weighting}

Recall from Theorem~\ref{thm:plpiw-11-12-21-22-is-np-hard} that
$\plpiw{MB}{SM}$ and $\plpiw{MB}{AM}$ 
are $\np$-complete.  We now show
that each of these problems is fixed-parameter tractable when
parameterized by the
budget.
To this end, recall the
\textsc{Knapsack} problem that was defined in the proof of
Theorem~\ref{thm:plpiw-11-12-21-22-is-np-hard}:
Given two finite lists of binary encoded integers,
$(c_i)_{i=1}^n$ (a list of costs) and $(p_i)_{i=1}^n$
(a list of profits) associated to a list $(o_i)_{i=1}^n$
of objects, as well as two further
integers, $C$ and $P$ (both encoded in binary),
the question is whether there is a subset
$J$ of $\{1,\dots,n\}$ such that
$\sum_{i\in J}c_i\leq C$ and $\sum_{i\in J}p_i\geq P$.
Thus, putting all objects from $\{o_j\mid j\in J\}$ into your backpack
does not violate your cost constraint $C$ but does satisfy your profit
demand~$P$.
\textsc{Knapsack} is an \np-hard problem that allows a pseudo-polynomial time 
algorithm. More precisely, this means that if
all cost and profit values are  given in unary, a polynomial-time algorithm
can be provided by using dynamic programming
(see~\cite{kel-pfe-pis:b:knapsack} for details).
This yields PTAS results both for the minimization version
$\MIN$-\textsc{Knapsack} (where the goal is to minimize the costs, subject
to the profit lower bound) and for the maximization version
$\MAX$-\textsc{Knapsack} (where the goal is to maximize the profits, subject
to the cost upper bound).

\begin{theorem}
\label{thm-issueweighted-FPT}
$\plpiw{MB}{SM}$ and $\plpiw{MB}{AM}$
(parameterized by the budget
or by the objective)
are in $\FPT$.
\end{theorem}
\begin{proofs}
Since
the unweighted variants of both problems are in~$\p$, we can compute
the amount 
 $d_j$
of dollars to be spent to win referendum $r_j$ in polynomial time in
both cases.  The interesting cases are the weighted ones.
We  re-interpret the given
$\plpiw{MB}{Y}$ instance, where
$\textsc{Y} \in \{\textsc{SM},\textsc{AM}\}$, as a
\textsc{Knapsack} instance.

In the  $\plpiw{MB}{Y}$ instance, 
every issue $r_j$ has an associated  cost $d_j$ and weight $w_j$.
The aim is to find a set of issues, i.e., a set $J\subseteq\{1,\dots,n\}$,
such that $\sum_{j\in J}d_j\leq B$ and $\sum_{j\in J}w_j\geq O$.
Consider $r_j$ as  an object $o_j$ in a \textsc{Knapsack} instance with
cost $c_j=d_j$ and profit $p_j=w_j$, with the bounds $C=B$ and $P=O$.
Then the $J\subseteq\{1,\dots,n\}$ that is a solution to the  $\plpiw{MB}{Y}$ instance is
also a solution to the
\textsc{Knapsack} instance, and vice-versa.  Furthermore,
the pseudo-polynomial
algorithm that solves
\textsc{Knapsack} in time $\Oh(n 2^{|B|})$, where
$|B|$ denotes the length of the encoding of~$B$, also solves $\plpiw{MB}{Y}$.

Observe that the given reduction works for both \textsc{SM} and \textsc{AM} because 
we introduce only one referendum.
We now use the pseudo-polynomial
algorithm to solve
\textsc{Knapsack} in time $\Oh(n 2^{|O|})$, where
$|O|$ denotes the length of the encoding of~$O$.~\end{proofs}

Voter bribery with issue weighting keeps its complexity status for both
evaluation criteria.

Since we can incorporate issue weights into brute-force
computations, we have the following corollary to
Theorems~\ref{thm:plp3j-FPT} and~\ref{thm:plp31-FPT}.

\begin{corollary}
\label{cor:vb-sm-plp-wiw-vb-am-plp-wiw-budget-and-disc-level}
\begin{enumerate}
\item 
  $\plpiw{VB}{SM}$ and $\plpiw{VB}{AM}$ (parameterized by the budget
  and by the discretization level) are in $\FPT$.
\item $\plpiw{VB}{SM}$ (parameterized by the budget) is in $\FPT$.
\end{enumerate}
\end{corollary}

It is not hard to transfer the $\wtwo$-hardness results from the
unweighted to the weighted case. However, it is unclear to us if 
or how the membership proofs of the preceding section transfer.
The difficulty appears to lie in the weights that the reducing machine or 
the produced Turing machine would have to handle.
Since it is  not known in advance which items will be bribed to
meet the objective requirement~$O$, the summation of item weights
cannot be performed by the reducing machine, but must be done by the
produced nondeterministic multi-tape Turing machine.
However, this Turing machine may only use time that can be measured in
the parameter, which has been budget per issue in the unweighted case.
We do not see how to do this.
 Therefore, we can state only the following.

\begin{corollary}
\label{cor:wiw-wtwo-hard}
  $\plpiw{VB}{SM}$ and $\plpiw{VB}{AM}$ (parameterized by the budget
  per issue)
 are $\wtwo$-hard.
\end{corollary}

 \begin{theorem}
 \label{thm:plpiw31}
 $\plpiw{VB}{SM}$ %
(parameterized by the budget per
 issue
 and by the objective)
 is in $\wtwo$.
 \end{theorem}
\begin{proofs}
 Membership in $\wtwo$ is a bit more tricky than in the unweighted case
 from Theorem~\ref{thm:plpw2-a}.  
To show membership in $\wtwo$, we reduce
$\plpiw{VB}{SM}$ to \textsc{SMNTMC}.
We describe how a nondeterministic multi-tape Turing
machine $M$ can solve such a lobbying problem.

Consider an instance of $\plpiw{VB}{SM}$: a probability matrix $P \in
\matrixrationalszeroone{m}{n}$ with a
cost matrix 
$C_P$,
a budget~$B$, and a fixed threshold~$t$, as well as an objective~$O$.
We may again identify $t$ with a
certain step level for the
price functions.

We describe the work of the reducing machine in the following.
 \begin{itemize}
  \item The reducing machine calculates the difference $O'$ between the
 target weight and the sum of the weights of the referenda that are
 already won.
 \item The reducing machine replaces issue weights bigger than $O'$ with
 $O'+1$. 
\item The reducing machine eliminates all referenda that are already won.
 \item For each referendum that is not already won, the reducing
 machine introduces a special letter $r_i$ to be used on tape~$T_0$ of the Turing machine to be constructed.
 \end{itemize}

For simplicity, we assume in the following that $O'=O$, i.e., the reduction
steps described above have not changed the instance. 

After these steps, all the weights are bounded by the second parameter. 
Moreover, recall that by the problem definition, the weight associated
to each issue is at least one.

Otherwise, the reducing machine works as described in the proof of Theorem~\ref{thm:plpw2-a}. In particular, from $P$, $C_P$, and $t$, the
machine extracts the information $H_{i,j}(d)$, $1\leq d \leq B$,
and then
 $s(j)$  (per $r_j$)
and finally the \emph{winning table} 
$W_i(\ell)$. %

We now describe how the Turing machine $M$ works.

 \begin{itemize}
 \item $M$
begins by guessing $\nu\leq O$ referenda that should also be won.
 (Recall that the weight associated to each issue is at least one, so guessing more than $O$ referenda is not necessary.)
 Then, $M$ spends $\Oh(f(O))$ time calculating
 whether winning those guessed referenda $r_{\gamma(1)},\dots,r_{\gamma(\nu)}$, $\nu\leq O$,
 would be sufficient to exceed the threshold~$O$. If not,
 $M$ halts and rejects, otherwise it continues working.
\item The mentioned threshold test can be implemented as follows:
$M$ guesses at most $O$ referenda and writes the corresponding
symbols on tape $T_0$. It first checks if all guessed referenda are pairwise different. If not, $M$ stops and rejects. Otherwise, 
$M$ writes a special symbol $O-1$ times on the second tape, $T_1$,
and moves its head on this tape to the right.
Then, it moves its head $w_j$ steps to the left on tape $T_1$ (replacing the special symbol by the blank symbol again) upon reading symbol $r_j$ on tape $T_0$. 
The threshold $O$ is passed if and only if the blank symbol is read
on $T_1$ after processing $r_{\gamma(\nu)}$ on tape $T_0$. 
Notice that tape $T_1$ will be empty again after this phase, while
tape $T_0$ will still contain the guessed referenda. 
Observe that the time needed by the Turing machine $M$ depends only on~$O$. 
 \item $M$ will then continue to work as described in
   the proof of Theorem~\ref{thm:plpw2-a}. During this phase, the weights
will be completely ignored. In particular, this part of $M$
can be produced in polynomial time by the reducing machine.
 As can be seen  in
   the proof of Theorem~\ref{thm:plpw2-a}, 
the time needed by $M$ in this phase depends only
on~$\nicefrac{B}{n}$.
\item %
Finally, $M$ will verify %
 if all (at most $O$) referenda guessed 
initially have been won.
In contrast to  the  proof of Theorem~\ref{thm:plpw2-a}, 
this test can now be implemented in a sequential fashion (upon reading $r_j$
on tape $T_0$, the corresponding test is performed on tape $T_j$),
needing time dependent on~$O$.
 \end{itemize}
 
This concludes the proof sketch.~\end{proofs}

It might be that $\wtwo$ is not the smallest class in the W-hierarchy
where the problem discussed in the preceding theorem could be placed.
However, we do not know how to find an FPT or $\wone$ algorithm for
it, even in the case when all weights equal one.  This is in contrast
to the possibly related problem {\sc Partial $t$-Domination}, which asks
whether there is a set of at most $k$ vertices in a graph that dominates at
least $t$ vertices. Our belief that these two problems
are related is motivated by
the fact that the classical dominating set problem
was the starting point
of the reduction showing hardness for {\sc Optimal Lobbying}.
Kneis, M\"olle, and Rossmanith showed that the problem
{\sc Partial $t$-Domination}
is in FPT even when parameterized by the threshold parameter
$t$ alone~\cite{KneMolRos2007}.

\section{Approximability}
\label{sec:approximability}

As seen in Tables~\ref{tab:plp} and~\ref{tab:plpiw}, many problem
variants of probabilistic lobbying are $\np$-complete.
Hence, it is interesting to
study them not only from the viewpoint of parameterized complexity,
but also from the viewpoint of approximability.

The budget constraint on the bribery problems studied so far gives
rise to natural minimization problems: Try to minimize the amount
spent on bribing.  For clarity, let us denote 
these minimization problems by prefixing the problem name with $\MIN$,
leading to, e.g., $\MIN$-$\ol$.

\subsection{Voter Bribery is Hard to Approximate}

The already-mentioned reduction of
Christian et al.~\cite{chr-fel-ros-sli:j:lobbying} (that proved that $\ol$ is
$\wtwo$-hard) is parameter-preserving (regarding the budget). The
reduction also\ has the property that a possible solution found in the $\ol$
instance can be re-interpreted as a solution to the
\textsc{Dominating Set} instance that the reduction started with, and the
$\ol$ solution and the \textsc{Dominating Set} solution are of the
same size. This in particular means that inapproximability results for
\textsc{Dominating Set} transfer to inapproximability results for
$\ol$. Similar observations are true for the interrelation of
\textsc{Set Cover} and \textsc{Dominating Set}, as well as for $\ol$
and $\plpiw{VB}{SM}$ (or $\plpiw{VB}{AM}$).

The known inapproximability
results~\cite{bel-gol-lun-rus:c:efficient-pcp-approximations,raz-saf:c:subconstant-lowdegree-test-pcp} for
\textsc{Set Cover} hence give the following result (see also
Footnote~4
in~\cite{san-sur-gil-lev:c:combinatorial-auction-generalizations}).

\begin{theorem} 
There is a constant $c>0$ such that
$\MIN$-$\ol$
is not approximable within factor
$c\cdot\log n$ unless $\np\subset\dtime(n^{\log\log n})$,
where $n$ denotes the number of issues.
\end{theorem}

Since $\ol$ can be viewed as a special case of both problem
$\plp{VB}{Y}$ and problem
$\plpiw{VB}{Y}$ for $\textsc{Y} \in \{\textsc{SM},\textsc{AM}\}$,
we have the following corollary.

\begin{corollary}
For $\textsc{Y} \in \{\textsc{SM},\textsc{AM}\}$,
there is a constant $c_{\textsc{Y}}>0$ such that 
both $\MIN$-$\plp{VB}{Y}$ and $\MIN$-$\plpiw{VB}{Y}$ are not
approximable within factor $c_{\textsc{Y}}\cdot \log n$ unless
$\np\subset\dtime(n^{\log\log n})$, where $n$ denotes the number of
issues.
\end{corollary}

\begin{proofs}
The proof of Theorem~\ref{thm:plpnp} explains in detail how to
interpret an instance of $\ol$ as a
$\plp{VB}{\textsc{Y}}$ instance, $\textsc{Y} \in
\{\textsc{SM},\textsc{AM}\}$.
The relation $B=n\cdot b$ between the budget $B$ and
the number of voters $b$ holds for both optimum and approximate solutions.
Hence, the $n$ is canceled
out when looking at the approximation ratio.~\end{proofs}

Conversely, a logarithmic-factor approximation can be given for the
minimum-budget versions of all our problems, as we will show now. We
first discuss the relation to the well-known \textsc{Set Cover}
problem, sketching a tempting, yet flawed reduction and pointing out
its pitfalls. Avoiding these pitfalls, we then give an
approximation algorithm for
$\MIN$-$\plp{VB}{AM}$.  Moreover, we define
the notion of cover number, which allows to state inapproximability
results for
$\MIN$-$\plp{VB}{AM}$.  Similar results hold for
$\MIN$-$\plp{VB}{SM}$, the constructions are sketched at the end of
the section.

Voter bribery problems are closely related to set cover problems, in
particular in the average-majority scenario, so that we should be able to carry over
approximability ideas from that area.  The intuitive translation of a
$\MIN$-$\plp{VB}{AM}$ instance
into a \textsc{Set Cover} instance is as follows: The universe of the derived
\textsc{Set Cover} instance should be the set of
issues, and the sets (in the \textsc{Set Cover} instance) are formed by considering the sets of
issues that could be influenced (by changing a voter's opinion)
through bribery of a specific voter.  Namely, when we pay voter $v$ a
specific amount of money, say $d$ dollars, he or she will credit 
$\nicefrac{d}{n}$ dollars to each issue and possibly change $v$'s opinion
(or at least raise $v$'s acceptance probability to some ``higher
level'').  The weights associated with the sets of issues correspond to
the bribery costs that are (minimally) incurred to lift the issues in
the set to some ``higher level.''
There are four differences to classical set covering problems:
\begin{enumerate}
\item Multiple voters may cover the same set of issues (with
different bribing costs).

\item The sets associated with one voter are not independent. For each voter,
the sets of issues that can be influenced by bribing that voter are linearly
ordered by set inclusion. Moreover, when bribing a specific voter, we have
  to first influence the ``smaller sets'' (which might be expensive)
  before possibly influencing the ``larger ones''; so, weights are
  attached to set differences, rather than to sets.
\item A \emph{cover number} $c(r_j)$ is associated with each
  issue~$r_j$, indicating by how many levels voters must raise their
  acceptance probabilities in order to arrive at average majority
  for~$r_j$.  The cover numbers can be computed beforehand
  for a given instance.  Then, we can also associate cover numbers with
  sets of issues (by summation), which finally leads to the cover
  number $N=\sum_{j=1}^n c(r_j)$ of the whole instance.
\item The money paid ``per issue'' might not have been sufficient for
  influencing a certain issue up to a certain level, but it is not
  ``lost''; rather, it would make the next bribery step cheaper, hence
  (again) changing weights in the
set cover interpretation. 
\end{enumerate}

To understand these connections better, let us have another look at
our running example (under
voter bribery with average-majority
evaluation, i.e.,
$\MIN$-$\plp{VB}{AM}$),
assuming an all-ones target vector.  If we paid $30$
dollars to voter~$v_1$, he or she would credit $10$ dollars to each
issue, which would raise his or her acceptance probability for the
second issue from $0.3$ to $0.4$; no other issue level is changed.
Hence, this would
correspond to a set containing only $r_2$ with weight $30$.  Note that
by this bribery, the costs for raising the acceptance probability of
voter $v_1$ to the next level would be lowered for the other two
issues. For example, spending 15 more dollars on $v_1$ would raise
$r_3$ from $0.5$ to $0.6$, since all in all $45$ dollars have been spent on
voter $v_1$, which means $15$ dollars per issue.  If the threshold is
$60\%$ in that example, then the first issue is already accepted (as
desired by The Lobby), but the second issue has gone up from $0.5$ to 
only $0.55$
on average, which means that we have to raise either the acceptance
probability of one voter by two levels (for example, by paying $210$
dollars to voter~$v_1$), or we have to raise the acceptance
probability of each voter by one level (by paying $30$ dollars to voter
$v_1$ and another $30$ dollars to voter~$v_2$).  This can be expressed
by saying that the first issue has a cover number of zero, and the
second has a cover number of two.

When we interpret an $\ol$ instance as a
$\plp{VB}{AM}$ instance, 
the cover number of the
resulting instance equals the number of issues, assuming
that the votes for all issues need amendment.
Thus we have the following corollary:

\begin{corollary}
There is a constant $c>0$ such that
$\MIN$-$\plp{VB}{AM}$ is not approximable within factor
$c\cdot \log N$ unless $\np\subset\dtime(N^{\log\log N})$, where
$N$ is the cover number of the given instance.
A fortiori, the same statement holds for
 $\MIN$-$\plpiw{VB}{AM}$. 
\end{corollary}

Let $H$ denote the harmonic sum function, i.e., $H(r)=\sum_{i=1}^r
\nicefrac{1}{i}$.  It is well known that $H(r)=O(\log r )$. More
precisely, it is known that
\[
\ln r 
 \leq H(r) \leq 
\ln r 
 +1.
\]

We now show the following theorem.

\begin{theorem}
\label{thm:plp-32-approximation}
$\MIN$-$\plp{VB}{AM}$
can be approximated within a factor of $\ln(N)+1$, where $N$ is the
cover number of the given instance.
\end{theorem}

\begin{proofs}
Consider the greedy algorithm shown in
Figure~\ref{fig:greedy-algorithm}, where $t$ is the given threshold
and we assume that The Lobby has the
all-ones target vector.  Note that the cover numbers (per referendum)
can be computed from the cost matrix $C_P$ and the threshold $t$
before calling the algorithm the very first time.

\begin{figure}[t!]
\center
\fbox{
\parbox{11.4cm}
{
\begin{description}
 \item[{\bf Input:}] A probability matrix $P \in
\matrixrationalszeroone{m}{n}$ (implicitly specifying a set $V$ of $m$
voters and a set $R$ of $n$ referenda),
a cost matrix~$C_P$, a treshold~$t$, and $n$ cover numbers
$c(r_1), \ldots , c(r_n) \in \nats$.
\end{description}
\begin{enumerate}
\item Delete referenda that are already won (indicated by $c(r_j)=0$),
  and modify $R$ and $C_P$ accordingly.
\item If $R=\emptyset$ then output the amount spent on bribing so far
   and STOP.
\item For each voter $v$, compute the least amount of money, $d_v$,
  that could raise any level in $C_P$. Let $n_v$ be the number of
  referenda whose levels are raised when spending $d_v$ dollars on
  voter~$v$.
\item
 Bribe voter $v$ such that $\nicefrac{d_v}{n_v}$ is minimum.
\item Modify $C_P$ by subtracting $\nicefrac{d_v}{n}$ from each amount
  listed for voter~$v$.
\item Modify $c$ by subtracting one from $c(r)$ for those referenda $r
  \in R$ influenced by this bribery action.
\item Recurse.
\end{enumerate}
}}
\caption{\label{fig:greedy-algorithm} Greedy approximation algorithm
for
$\MIN$-$\plp{VB}{AM}$ in Theorem~\ref{thm:plp-32-approximation}
}
\end{figure}

Observe that our greedy algorithm influences voters by raising
their acceptance probabilities by only one level, so that the amount
$d_v$ possibly spent on voter $v$ in Step~3 of the algorithm actually
corresponds to a set of referenda; we do not have to consider
multiplicities of issues (raised over several levels) here.

Let $S_1,\dots,S_\ell$ be the sequence of sets of referenda picked by
the greedy bribery algorithm, along with the sequence
$v_1,\dots,v_\ell$ of voters and the sequence $d_1,\dots, d_\ell$ of
bribery dollars spent this way.  Let
$R_1=R,\dots,R_\ell,R_{\ell+1}=\emptyset$ be the corresponding
sequence of sets of referenda, with the accordingly modified cover
numbers $c_i$.  Let $j(r,k)$ denote the index of the set in the
sequence influencing referendum $r$ the $k$th time with $k\leq c(r)$,
i.e., $r\in S_{j(r,k)}$ and $|\{i<j(r,k)\condition r\in S_i\}|=k-1$.
To cover $r$ the $k$th time, we have to pay
$\chi(r,k)=\nicefrac{d_{j(r,k)}}{|S_{j(r,k)}|}$ dollars.  The greedy
algorithm will incur a cost of $\chi_{\textit{greedy}}=\sum_{r\in
  R}\sum_{k=1}^{c(r)} \chi(r,k)$ in total.

An alternative view of the greedy algorithm is from the perspective of
the referenda: By running the algorithm, we implicitly define a
sequence $s_1,\dots,s_N$ of referenda, where $N=c(R)=\sum_{r\in
R}c(r)$ is the cover number of the original instance, such that
$S_1=\{s_{\lambda(1)},\dots,s_{\rho(1)}\}$,
$S_2=\{s_{\lambda(2)},\dots,s_{\rho(2)}\}$,
\ldots,
$S_{\ell}=\{s_{\lambda(\ell)},\dots,s_{\rho(\ell)}\}$,
where $\lambda, \rho:\{1,\dots,\ell\}\to \{1,\dots,N\}$ are functions
such that 
$\lambda(i)$ gives the element of $S_i$ with the smallest subscript and
$\rho(i)$ gives the element of $S_i$ with the greatest subscript for
each~$i$, $1 \leq i \leq \ell$:
\begin{eqnarray*}
\lambda(i) = 1 + \sum_{j<i}|S_j| & \text{ and } & 
\rho(i)    = \sum_{j \leq i}|S_j|.
\end{eqnarray*}
Ties (how to list
elements within any~$S_i$) are broken arbitrarily.  Slightly abusing 
notation,
we associate a cost $\chi'(s_i)$ with $S_i$ for each $i$,
(keeping in mind the multiplicities of covering implied by the
sequence $\langle S_i \rangle_i$), so that 
$\chi_{\textit{greedy}}=\sum_{i=1}^N\chi'(s_i)$.
Note that $d_i=\sum_{\lambda(i)\leq j\leq \rho(i)}\chi'(s_j)$.

Consider $s_j$ with $\lambda(i)\leq j\leq \rho(i)$. The current referendum set
$R_{i}$ has cover number $N-\lambda(i)+1$, i.e., of at least $N-j+1$.  Let
$\chi_{\textit{opt}}$ be the cost of an optimum bribery strategy
$\mathcal{C}^*$ of the original universe.  $\mathcal{C}^*$ also yields
a cover of the referendum set $R_{i}$ with cost at most
$\chi_{\textit{opt}}$. The average cost per element (taking into
account multiplicities as given by the cover numbers) is
$\nicefrac{\chi_{\textit{opt}}}{c(R_i)}$.  (So, whether or not some
new levels are obtained through bribery does not really matter here.)
$\mathcal{C}^*$ can be described by a sequence of sets of referenda
$C_1,\dots,C_q$, with corresponding voters $z_1,\dots,z_q$ and
dollars $d_1^*,\dots, d_q^*$ spent.  Hence,
$\chi_{\textit{opt}}=\sum_{\kappa=1}^qd_\kappa^*$.
With each bribery step we associate the cost factor
$\nicefrac{d_\kappa^*}{|C_\kappa|}$, for each issue
$r$ contained in $C_\kappa$.
$\mathcal{C}^*$ could be also viewed as a bribery strategy for
$R_i$.  By the pigeon hole principle, there is a referendum $r$ in $R_i$ (to 
be
influenced the $k$th time) with cost factor at most
$\nicefrac{d_\kappa^*}{|C_\kappa\cap R_i|}\leq
\nicefrac{\chi_{\textit{opt}}}{c(R_i)}$, where $\kappa$ is the index
such that $C_\kappa$ contains $r$ for the $k$th time in
$\mathcal{C}^*$ (usually, the cost would be smaller, since part of the
bribery has already been paid before).  Since $(S_i,v_i)$ was picked
so as to minimize $\nicefrac{d_i}{|S_i|}$, we find
$\nicefrac{d_i}{|S_i|}\leq \nicefrac{d_\kappa^*}{|C_\kappa\cap
R_i|}\leq \nicefrac{\chi_{\textit{opt}}}{c(R_i)}$.

We conclude that
\[
\chi'(s_j)\leq
\frac{\chi_{\textit{opt}}}{c(R_i)}=
\frac{\chi_{\textit{opt}}}{N-\lambda(i)+1}\leq
\frac{\chi_{\textit{opt}}}{N-j+1}.
\]
Hence,
\[
\chi_{\textit{greedy}}
  =  \sum_{j=1}^N\chi'(s_j)
 \leq \sum_{j=1}^N\frac{\chi_{\textit{opt}}}{N-j+1}
  =  H(N) \chi_{\textit{opt}}
 \leq (\ln(N)+1)\chi_{\textit{opt}},
\]
which completes the proof.~\end{proofs}

In the strict-majority scenario (\textsc{SM}),
cover numbers would have a different
meaning---we thus call them \emph{strict cover numbers}: For each
referendum, the corresponding strict cover number tells in advance how
many voters
have to change their opinions (bringing them individually over the
given threshold~$t$) to accept this referendum.  Again, the strict
cover number of a problem instance is the sum of the strict cover
numbers of all given referenda.

The corresponding
greedy algorithm would therefore choose to influence voter $v_i$ (with
$d_i$ dollars) in
the $i$th loop so that $v_i$ changes his or her opinion on some
referendum $r_j$ such that $\nicefrac{d_i}{|\rho_j|}$ is
minimized.\footnote{Possibly, there is a whole set
$\rho_j$ 
of referenda influenced this way.}

We can now read the approximation bound proof given for the
average-majority scenario nearly literally as before, by re-interpreting the
formulation ``influencing referendum $r$'' meaning now a complete
change of opinion for a certain voter (not just gaining one level
somehow).  This establishes the following result.

\begin{theorem}
\label{thm:plp-31-approximation}
$\MIN$-$\plp{VB}{SM}$
can be approximated within a factor of $\ln(N)+1$, where
$N$ is the strict cover number of the given instance.
\end{theorem}

Note that this result is in some sense stronger than
Theorem~\ref{thm:plp-32-approximation} (which refers to the
average-majority scenario), since the cover number of an instance
could be larger than the strict cover number.

This approximation result is complemented by a corresponding hardness
result.

\begin{theorem}
There is a constant $c>0$ such that %
$\MIN$-$\plp{VB}{SM}$ is not approximable within factor
$c\cdot \log N$ unless $\np\subset\dtime(N^{\log\log N})$, where
$N$ is the strict cover number of the given instance.
A fortiori, the same statement holds for $\MIN$-$\plpiw{VB}{SM}$.
\end{theorem}

Unfortunately, those greedy algorithms do not (immediately) transfer
to the case when issue weights are allowed. These weights
might also influence the quality of approximation, but a simplistic
greedy algorithm might result in covering the ``wrong'' issues.
Also, the proof of the approximation factor given above will not carry
over, since we need as one of the proof's basic ingredients that an
optimum solution can be interpreted as a partial one at some point.
Those problems tend to have a different
flavor.

\subsection{Polynomial-Time Approximation Schemes}

Those problems for which we obtained $\FPT$ results in the case of
issue weights actually enjoy a polynomial-time approximation scheme
(PTAS) when viewed as minimization
problems.\footnote{A polynomial-time approximation scheme is an
algorithm that for each pair $(x,\epsilon)$, where $x$
is an instance of an optimization problem and $\epsilon > 0$ is a
rational constant, runs in time polynomial in $|x|$ (for each fixed value
of $\epsilon$) and outputs an
approximate solution for $x$ within a factor of~$1+\epsilon$.}
The proof of Theorem~\ref{thm-issueweighted-FPT} yields a {PTAS}.
That result
was obtained by transferring pseudo-polynomial time algorithms:
For each fixed value of the parameter (the budget $B$ or the objective~$O$),
we obtain a polynomial-time algorithm for the decision problem, which
can be used to approximate the minimization problem.

\begin{theorem}
\label{thm-min-issueweighted-PTAS}
Both $\MIN$-$\plpiw{MB}{SM}$ and $\MIN$-$\plpiw{MB}{AM}$
admit a PTAS.
\end{theorem}

\begin{proofs}We provide some of the details for $\MIN$-$\plpiw{MB}{SM}$ only.
As in the  proof of Theorem~\ref{thm-issueweighted-FPT}, we first compute
the amount, $d_j$,
 to be spent to win referendum $r_j$ in polynomial time.
We then re-interpret the given
instance of $\MIN$-$\plpiw{MB}{SM}$  
as a
$\MIN$-\textsc{Knapsack} instance.
After
this re-interpretation, 
every issue $r_j$ has an associated cost $d_j$ and weight~$w_j$.
The aim is to find a set of issues, i.e., a set $J\subseteq\{1,\dots,n\}$,
such that  $\sum_{j\in J}w_j\geq O$ and $\sum_{j\in J}d_j$ is minimum.
Consider $r_j$ as  an object $o_j$ in a $\MIN$-\textsc{Knapsack} instance with
cost $c_j=d_j$ and profit $p_j=w_j$, with the bound  $P=O$.
Then the subset $J\subseteq\{1,\dots,n\}$
that is a solution to the  $\plpiw{MB}{SM}$ instance is
also a solution to the
\textsc{Knapsack} instance, and vice-versa.  Furthermore,
the PTAS 
algorithm that approximates 
$\MIN$-\textsc{Knapsack}  also gives a PTAS for  $\plpiw{MB}{SM}$.~\end{proofs}

Let us mention here that the issue-weighted problem variants
are actually bicriteria problems: We want to achieve as much as possible
(expressed by the objective $O$) and pay as little as possible (expressed by the budget $B$).
So, we could also consider this as a maximization problem (where now $B$ becomes
again part of the input).
By the close relation to \textsc{Knapsack} mentioned above, the pseudo-polynomial time
algorithms again result in PTAS results for this model:

\begin{theorem}
\label{thm-max-issueweighted-PTAS}
Both $\MAX$-$\plpiw{MB}{SM}$ and $\MAX$-$\plpiw{MB}{AM}$
admit a PTAS.
\end{theorem}

It would be interesting to study this optimization criterion in the light of other bribery scenarios.
\section{Conclusions}
\label{sec:conclusions}

This paper lies at the intersection of three research streams:
computational social choice, reasoning under uncertainty, and computational
complexity.  It lays the foundation for further study of vote influencing 
in stochastic settings.  We have shown that 
uncertainty complicates the picture; the choice of model strongly 
affects computational complexity.

We have studied four lobbying scenarios in a probabilistic
setting, both with and without issue weights.  Among these, we identified
problems that can be solved in polynomial
time, problems that are $\np$-complete yet fixed-parameter tractable, and
problems that are hard (namely, $\wtwo$-complete or $\wtwo$-hard)
in terms of their parameterized complexity with suitable parameters.  
We also investigated the approximability of hard probabilistic
lobbying problems (without issue weights) and obtained both
approximation and inapproximability results.

An interesting direction for future work would be to study the
parameterized complexity of such problems
under different parameterizations.  
We would also like to investigate the open
question of whether one can find
logarithmic-factor approximations for voter bribery with issue 
weights.  

From the viewpoint of parameterized complexity, 
it would be interesting to solve the questions left open in this paper
(see the question marks in Tables~\ref{tab:plp} and~\ref{tab:plpiw}),
in particular regarding voter bribery under average majority, with and
without issue weighting, parameterized by the budget or by the budget
per issue.
The parameterized complexity of the 
problems involving microbribery with issue weighting,
under either strict majority or average majority,
parameterized by the budget per issue, is an open 
issue as well (again, see Table~\ref{tab:plpiw}).
In fact, we did not see how to put these problems
in any level of the W-hierarchy. Problems that
involve numbers seem to have a particular
flavor that makes them hard to tackle with these
techniques~\cite{fel-gas-ros:c:parameterizing-by-the-number-of-numbers},
but we suggest this as the subject of further
studies.
Note that all probabilistic lobbying 
problems that have been connected with $\wtwo$ somehow in this paper
are in fact strongly $\np$-hard, i.e.,
their hardness does not depend on whether the
numbers in their inputs are encoded in unary or in binary. We suspect that this
adds to the difficulty when dealing with these problems from a parameterized perspective.

{\small
\bibliographystyle{alpha}
\bibliography{lobbying}
} %

\end{document}